\documentclass{jfm}

\usepackage{multirow}%
\usepackage{mathrsfs}%
\usepackage[title]{appendix}%
\usepackage[dvipsnames]{xcolor}%
\usepackage{textcomp}%
\usepackage{manyfoot}%
\usepackage{booktabs}%
\usepackage{algorithm}%
\usepackage{algorithmicx}%
\usepackage{algpseudocode}%
\usepackage{listings}%
\usepackage{subcaption}
\usepackage[justification=justified,format=plain]{caption}
\usepackage{placeins}
\usepackage[normalem]{ulem}
\usepackage{xpatch}
\usepackage{graphicx}
\usepackage{epstopdf}
\usepackage{newtxtext}
\usepackage{newtxmath}
\usepackage{natbib}
\usepackage{hyperref}
\usepackage{amsmath}
\usepackage{etoolbox}
\hypersetup{
    colorlinks = true,
    urlcolor   = blue,
    citecolor  = black,
}

\newcommand{\RomanNumeralCaps}[1]

\shorttitle{Interaction between a bubble and a vortex ring}
\shortauthor{C. Estepa-Cantero et al.}

\title{Three-dimensional experimental investigation of the interaction between a rising bubble and a vortex ring}

\author{C. Estepa-Cantero\aff{1,2}, M. Lorite-D\'iez\aff{1,2}, J. Ruiz-Rus\aff{2,3}, S. Cazin\aff{4}, R. Bolaños-Jiménez\aff{2,3} \corresp{\email{rbolanos@ujaen.es}},
  P. Ern\aff{4}
 \and C. Martínez-Bazán\aff{1,2}.}
\affiliation{\aff{1}Área de Mecánica de Fluidos, Departamento de Mec\'anica de Estructuras e Ingenier\'{\i}a Hidr\'aulica, Universidad de Granada. Campus Fuentenueva s/n, 18071, Granada, Spain
\aff{2}Instituto Interuniversitario de Investigación del Sistema Tierra en Andalucía (IISTA), Universidades de Granada, Jaén y Córdoba. Avda. del Mediterr\'aneo s/n, 18006, Granada, Spain
\aff{3}Área de Mecánica de Fluidos, Departamento de Ingenier\'{\i}a Mec\'anica y Minera, Universidad de Jaén. Campus de las Lagunillas s/n, 23071, Jaén, Spain
\aff{4}Institut de M\'ecanique des Fluides de Toulouse (IMFT), CNRS, Toulouse INP and Universit\'e de Toulouse, 2 All. du Professeur Camille Soula, 31400 Toulouse, France
}

\begin{document}
\maketitle
\begin{abstract}
The interaction between turbulent flows and bubbles is a complex phenomenon ubiquitous in natural and industrial settings. In this work, we experimentally investigate, from a fundamental perspective, the interaction between a rising bubble and a vortex ring in counterflow. Using time-resolved three-dimensional Lagrangian Particle Tracking (4D-LPT) coupled with shadowgraphy, we obtain simultaneous measurements of the bubble motion and the surrounding liquid flow. This approach enables detailed observation of bubble dynamics, deformation, and eventual breakup, as well as the fluid motion. We examine several flow configurations by varying the vortex circulation and the Weber number while maintaining a comparable vortex-to-bubble size ratio. Based on these measurements, we classify the interaction events into three categories according to their impact on bubble dynamics and vortex stability over time. Through experiments, we address for the first time the three-dimensional effects of these interactions, which had not been considered in previous studies. The analysed experiments comprise: Case I, corresponding to a weak interaction in which neither the bubble nor the vortex is significantly affected; Case II, where the bubble is captured and advected by the vortex, leading to a strong distortion of the vortex due to the presence of the bubble within its core; and Case III, involving a stronger vortex capable of capturing the bubble and breaking it into two fragments without a severe loss of energy in the vortex core. The analysis of these results provides insight into the bubble breakup process and the mechanisms responsible for the destabilisation of the vortex ring.
\end{abstract}


\section{Introduction}\label{sec1}
Investigating the complex interactions between bubbles, inertial particles, and vortical structures present in turbulent flows is essential to understand a wide range of engineering operations and natural phenomena, including turbulent flows in chemical reactors, mixing in oceanic layers, ship hydrodynamics, and drag reduction applications~\citep{Berkman1988,Prince1990,Melville1996,Deane2002,van2013importance,ferrante2004effects}. In such systems, the presence and deformation of bubbles substantially influence the transport of mass, momentum, and energy within the continuous phase, while the turbulent flow exerts dynamic forces that dictate both bubble dynamics and the resulting size distribution. Despite their importance, the mechanisms by which the dispersed phase modulates turbulence, and how this modulation depends on key physical parameters, are poorly understood \citep{Ruetsch1994,Climent2006,Balachandar2010,Spandan2017}. This knowledge gap persists largely due to the intricate, non-linear nature of these two-way interactions, making them a subject of ongoing and fundamental research~\citep{Perrard2021,Pandey2021,Riviere2023}.

The bubble-turbulent flow interaction could lead to bubble breakup. In particular, bubble fragmentation in turbulent flows has been extensively studied since the foundational works of \citet{kolmogorov1949breakage} and \citet{hinze1955fundamentals}. These seminal studies established that bubbles or droplets break when turbulent stresses overcome surface tension forces \citep{martinez1999breakup,martinez1999daughter}, broadly categorised into inertial mechanisms at high Weber numbers, and resonance mechanisms at low Weber numbers~\citep{risso1998,revuelta2006breakup,revuelta2008breakup}. Here, the turbulent Weber number, $We_t= \rho \,\varepsilon^{2/3} R_{b}^{5/3}/\sigma$, is defined as the ratio between the turbulent stresses acting on the surface of the bubble, $\tau_{t} \sim \rho \,(\varepsilon R_{b})^{2/3}$, and the confining stresses due to surface tension, $\tau_s \sim \sigma/R_{b}$, where $\rho$ is the liquid density, $\sigma$ the surface tension, $\varepsilon$ the dissipation rate of turbulent kinetic energy per unit mass and $R_b=\left(3 \mathcal{V}_b/4\pi\right)^{1/3}$ is the bubble volumetric equivalent radius, with $\mathcal{V}_b$ being the bubble volume. A widely accepted hypothesis from this framework, known as the Kolmogorov-Hinze (KH) paradigm, postulates that bubbles are primarily broken by turbulent eddies of similar size~\citep{hinze1955fundamentals,Moin2020,masuk2021,qi2022fragmentation}. Larger eddies are believed to transport the bubble without breaking it, while smaller ones lack sufficient energy for rupture. However, testing this hypothesis has been challenging directly due to the multiscale nature of turbulence, where eddies of various sizes are present simultaneously \citep{qi2024breaking}.

To simplify the complexity of the interaction of bubbles and turbulent flows and gain fundamental insight, the interaction between a single bubble and a vortex ring, which serves as a model for a turbulent eddy, is an idealised scenario~\citep{chahine1995bubble,jha2015interaction,martinez2015bubbles}. In this case, the Weber number is defined as the ratio of the inertial forces associated with the vortex ring to the surface tension forces acting on the rising bubble, as it will be detailed later in Section \ref{sec:method}. This approach provides a controlled framework to unravel the underlying mechanisms governing bubble or particle deformation dynamics and their reciprocal influence on vortex behaviour.

A series of experimental studies has shed light on the complex interaction between vortex rings and bubbles or particles. \citet{sridhar1999effect} reported that microbubbles could distort a vortex ring core, with the core eventually regaining its initial state after the bubbles escaped. \citet{jha2015interaction} conducted controlled experiments in a co-current configuration by generating vortex rings and injecting air bubbles of known volume by their side, revealing a four-stage interaction process: capture, azimuthal elongation and gradual breakup, post-breakup motion, and eventual escape. Their work highlighted that bubbles could significantly distort vortex rings, particularly at low Weber numbers, exhibiting similarities with phenomena observed in turbulent bubbly flows, such as enstrophy reduction and distortion of structures. \citet{zednikova2019experiments} experimentally investigated bubble breakup triggered by vortex ring impact, establishing this interaction as a deterministic alternative to the stochastic nature of bubble fragmentation in turbulent flows. They quantified breakup frequency, the mean number of daughter bubbles, and size distributions using high-speed imaging. Building on this work, \citet{biswas2022interaction} carried out a series of experiments emphasising the role of deformability and size. A comparison between a deformable air bubble and a rigid buoyant particle demonstrated that bubble deformability fundamentally alters the interaction dynamics from the capture phase onward, with rigid particles causing a stronger disruption of the vortex structure. Moreover, the same authors focused on the bubble-to-vortex-core size ratio, finding that larger ratios at low $We$ induce stronger vortex core deformation and fragmentation, leading to marked reductions in both convection speed and azimuthal enstrophy~\citep{biswas2023vortex} (hereafter, the contribution of the azimuthal vorticity to the enstrophy, defined in equation~\ref{eq:eqs_integral}, will referred to as the azimuthal enstrophy, $\Omega_\theta$). They identified a scaling law for these reductions and analysed bubble elongation, breakup modes, and cascade breakup time. Most recently, ~\citet{biswas2025interaction} extended their analysis to rigid particles interacting with vortex rings while varying the particle-to-core size ratio. Results confirmed that larger particles significantly disrupt vortex dynamics, with effects more pronounced than those observed for deformable bubbles.

Nevertheless, most of these experimental studies rely on two-dimensional measurements or partial data. In this regard, numerical studies can complement them by providing detailed three-di\-men\-sional insight into flow fields and grant access to parameters that are challenging to measure experimentally. However, addressing the interaction of a vortex ring with a rising bubble is also complex numerically due to either the enormous computational cost of fully-resolved DNS or the need for experimental validation of simulation results based on partial modelling of the problem (such as turbulence, interface, among others). Early numerical studies by \citet{higuera2004axisymmetric} and \citet{revuelta2010interaction} explored the axisymmetric interaction of an inviscid vortex ring with a bubble, revealing different breakup modes depending on the vortex-to-bubble size ratio and Weber number. These studies primarily focused on bubble deformation and breakup occurring outside the vortex core. Other numerical investigations, such as those by \citet{chahine1995bubble} and \citet{ferrante2007effects}, explored scenarios where bubbles were entrained into vortex cores, observing extreme elongation, wrapping, and reduction of vorticity and enstrophy. \citet{foronda2021deformation} carried out fully three-di\-men\-sional simulations, systematically varying $We$ and the vortex-to-bubble size ratio. Their results, supported by experimental validation, revealed similar qualitative behaviour but suggested that the bubble presence does not significantly affect the global vortex enstrophy. Instead, they observed a redistribution of the enstrophy from the azimuthal to the radial and axial components following the loss of axial symmetry. They also highlighted that the sensitivity to initial conditions can lead to a probabilistic nature of bubble breakup, even in nominally deterministic systems. Similarly, \citet{cihonski2013volume} showed that the reaction forces exerted by bubbles on the fluid, under traditional two-way coupling assumptions, may be too small to significantly distort the vortex core. More recently, \citet{new2024collisions} used large-eddy simulations (LES) to study vortex ring collisions with hemispherical solids. Although focused on rigid bodies, their work provides valuable insight into the three-di\-men\-sional vortex dynamics and the formation of secondary and tertiary vortex structures during complex interactions. Together, these numerical studies demonstrate that high-fidelity simulations are essential to capture the fine-scale features of bubble–vortex and particle–vortex interactions, particularly those that elude direct measurement. However, while numerical simulations can provide a powerful theoretical framework, their inherent simplifications require the use of three-dimensional experimental techniques for validation.

Despite significant progress in understanding this problem, several key questions remain open. For instance, the exact mechanisms of turbulence modulation by bubbles, as well as their parametric dependence, are still not fully understood~\citep{jha2015interaction,biswas2022interaction}. Furthermore, the precise origins of vortex instabilities and core fragmentation at low $We$, and their role in the observed reduction of convection speed, remain unclear~\citep{jha2015interaction,martinez2015bubbles}. Discrepancies between experimental and numerical results regarding the redistribution of vorticity, enstrophy, and circulation also persist. Finally, the optimal vortex-to-bubble size ratio and the influence of the vortex core-to-vortex radius ratio ($a_0/R_0$) on breakup efficiency require further investigation, particularly in relation to the observed minimum critical Weber number~\citep{foronda2021deformation}.
 
Thus, the present work aims to address these open questions by examining the temporal and three-dimensional spatial evolution of the flow variables, focusing on three representative cases of bubble--vortex interactions: weak interaction; bubble drag and release, and bubble breakup. To achieve that, we will experimentally characterise the dynamics of the interaction of a bubble with a vortex ring using three-dimensional Lagrangian Particle Tracking (4D-LPT) coupled with tomographic reconstruction. Recent studies have demonstrated the capability of these three-di\-men\-sional techniques to simultaneously resolve the fluid velocity field and reconstruct the instantaneous body morphology, yielding valuable insights into the coupling between body kinematics, interface deformation, and wake dynamics for single freely rising bubbles \citep{pavlov2021exploration,chang2024,kameke2026} and solid bodies \citep{lorite2022experimental,giurgiu2023tu,caridi2025complete}. Building upon these methodological advances, the present experimental approach tackles situations involving complex fluid dynamics associated with strong bubble shape deformations, providing joint measurements of the three-di\-men\-sional carrier-phase flow field and of the dispersed-phase motion and shape. In this way, we can address the existing discrepancies in literature, provide a valuable experimental dataset for numerical simulations validation and unravel existing challenges, such as (i) the true three-dimensional evolution of vorticity and enstrophy during the interaction, allowing for definitive conclusions on whether vorticity is redistributed or dissipated; (ii) the complex three-dimensional nature of vortex instabilities and their development in the presence of a bubble; (iii) the coupling between bubble deformation and the three-dimensional flow field changes within the vortex core.

The work is organised as follows. Section \ref{sec:method} clarifies the experimental methodology, where we briefly present the experimental facility, as well as the different acquisition and analysis techniques used to describe the flow and bubble dynamics. More details on the methodology may be found in Appendices \ref{app:calib} to \ref{app:processing}.
Section \ref{sec:results} presents the results: first, the different types of interaction explored in this study are outlined in Section \ref{sec:Intro_cases}, followed by Sections \ref{subsec:caseI} to \ref{subsec:caseIII}, which describe the behaviour of Cases I to III, respectively. Finally, Section \ref{sec:conclusions} presents the conclusions and discussion of the results.

\section{Methodology}\label{sec:method}

\subsection{Description of the experiment}\label{subsec:set-up}

\begin{figure}
    \centering
    \includegraphics[width=.85\linewidth]{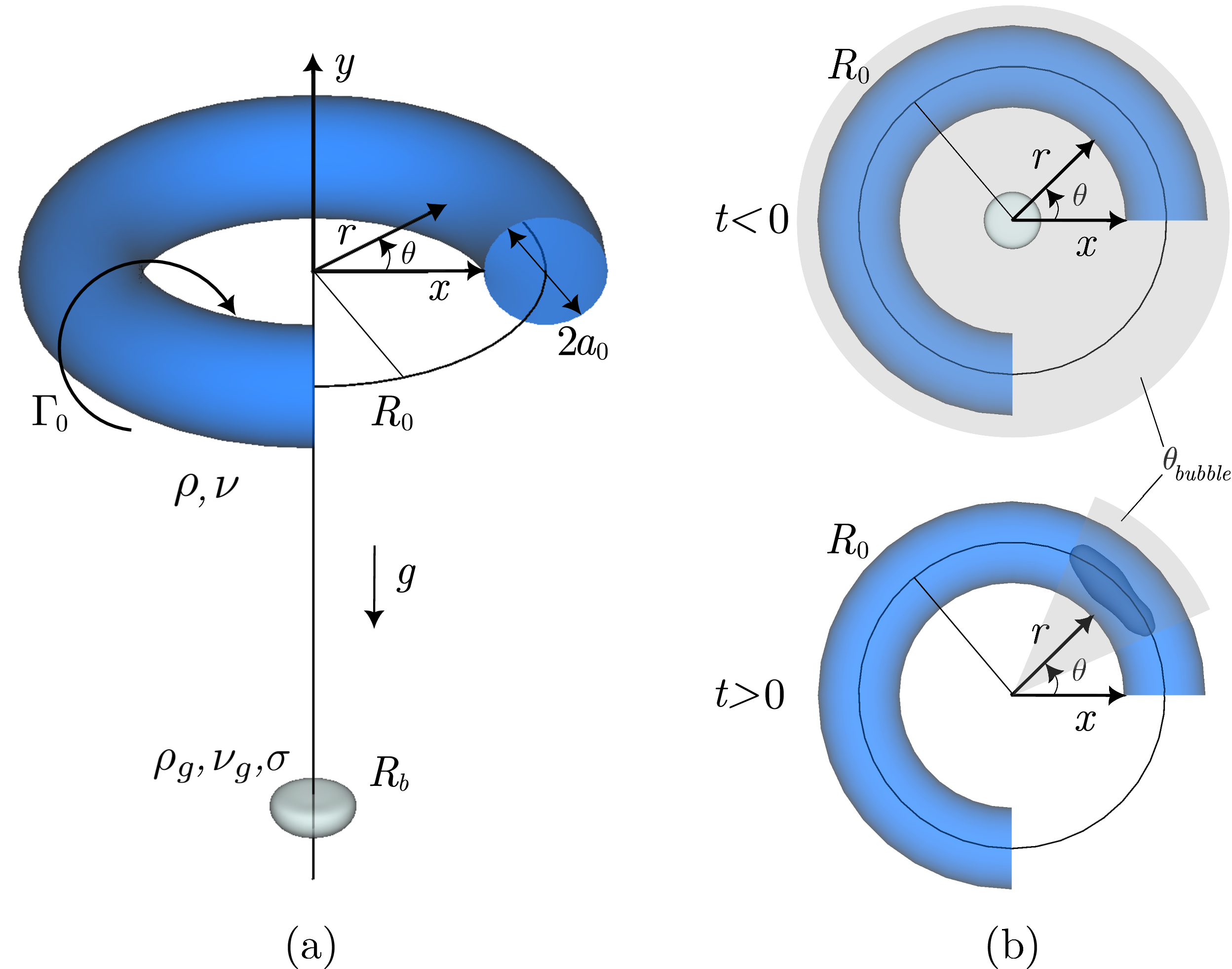}
    \vspace{1ex}
  \caption{(a) Three-dimensional sketch of the problem with the main involved variables and reference frames. (b) Top-view showing the azimuthal sector covered by the bubble ($\theta_{bubble}$, grey area) before (top) and during (bottom) the interaction.}
    \label{fig:problem_sketch}
\end{figure}

We consider a simplified configuration (see figure~\ref{fig:problem_sketch}) involving a single bubble of volumetric equivalent radius $R_b$, density $\rho_g$, kinematic viscosity $\nu_g$ and surface tension $\sigma$, rising in a still liquid with known density and kinematic viscosity, $\rho$ and $\nu$, respectively. When rising along the vertical direction (against gravity), the bubble interacts with a downwards-directed vortex ring of radius $R(t)$, core radius $a(t)$, and circulation $\Gamma(t)$, with their initial values denoted as $R_0$, $a_0$, and $\Gamma_0$ (see figure~\ref{fig:problem_sketch}a). These variables are obtained at $t=t_0$, when the vortex has completely entered the measurement volume. Here, $t_0<0$ since we define $t=0$ as the time when the bubble begins to interact with the vortex ring. The non-dimensional quantities are then computed using,
$R_0$, $R_0^2/\Gamma_0$ and $\Gamma_0/R_0$, as the characteristic length, time, and velocity, respectively. Note that we define the dimensionless circulation as $\Gamma_0/(U_0 R_0)$, based on the theoretical model for a Lamb-Oseen vortex ring, $U_0R_0=\Gamma_0/(4\pi)[\ln(8R_0/a_0)-1/4]$ \citep{lamb1924hydrodynamics}.

To describe the vortex-bubble interaction, two coordinate systems are used: a Cartesian system in which $y$ is the vertical direction and $x, z$ are the horizontal coordinates, and a cylindrical one in which the radial coordinate is $r = \sqrt{x^2 + z^2}$ and the azimuthal coordinate is $\theta$, with $y$ the vertical one (see figure~\ref{fig:problem_sketch}). Figure \ref{fig:problem_sketch}(b) shows a top view of the bubble-vortex interaction. Here, $\theta_{bubble}$ defines the range of azimuthal planes covered by the bubble (shaded area in figure \ref{fig:problem_sketch}b). Note that this value is affected by the bubble shape but also by its radial position: when the bubble is released ($t<0$, figure \ref{fig:problem_sketch}b top), it is nearly centred in the vertical axis and $\theta_{bubble}\approx 360^\circ$, but when the interaction takes place, the bubble moves radially towards the vortex core and, therefore, $\theta_{bubble}$ decreases.

Since $\rho \gg \rho_g$ and $\nu \gg \nu_g$, the vortex-bubble interaction can be described in terms of four dimensionless numbers: the Weber number, $We=\rho \, (\Gamma_0/2\pi R_0)^2/(\sigma/R_b)$, Reynolds number, $Re=\Gamma_0/\nu$, vortex-to-bubble size ratio ($R_0/R_b$), and core-to-vortex ring radius ratio ($a_0/R_0$). Note that $We$ and $Re$ are defined using $\Gamma_0/(2\pi R_0)$ as the characteristic velocity, with $\Gamma_0$ the initial vortex circulation. The bubble motion prior to its interaction with the vortex can be characterised in terms of the Bond, $Bo=\rho g R_b^2/\sigma$, and Galilei, $Ga=\sqrt{g R_b} \, R_b/\nu$, numbers. The combined values of these two parameters define the bubble rise regime and, in particular, determine its terminal velocity, $U_t$, or equivalently, its Froude number, $Fr = U_t/\sqrt{g R_b}$~\citep{cano2016paths}. Therefore, the bubble's Reynolds and Weber numbers can be expressed as $Re_b = Ga \, Fr$ and $We_b=Bo\,Fr^2$, respectively.

\begin{figure}
    \begin{subfigure}{.47\textwidth}
        \includegraphics[width=\linewidth]{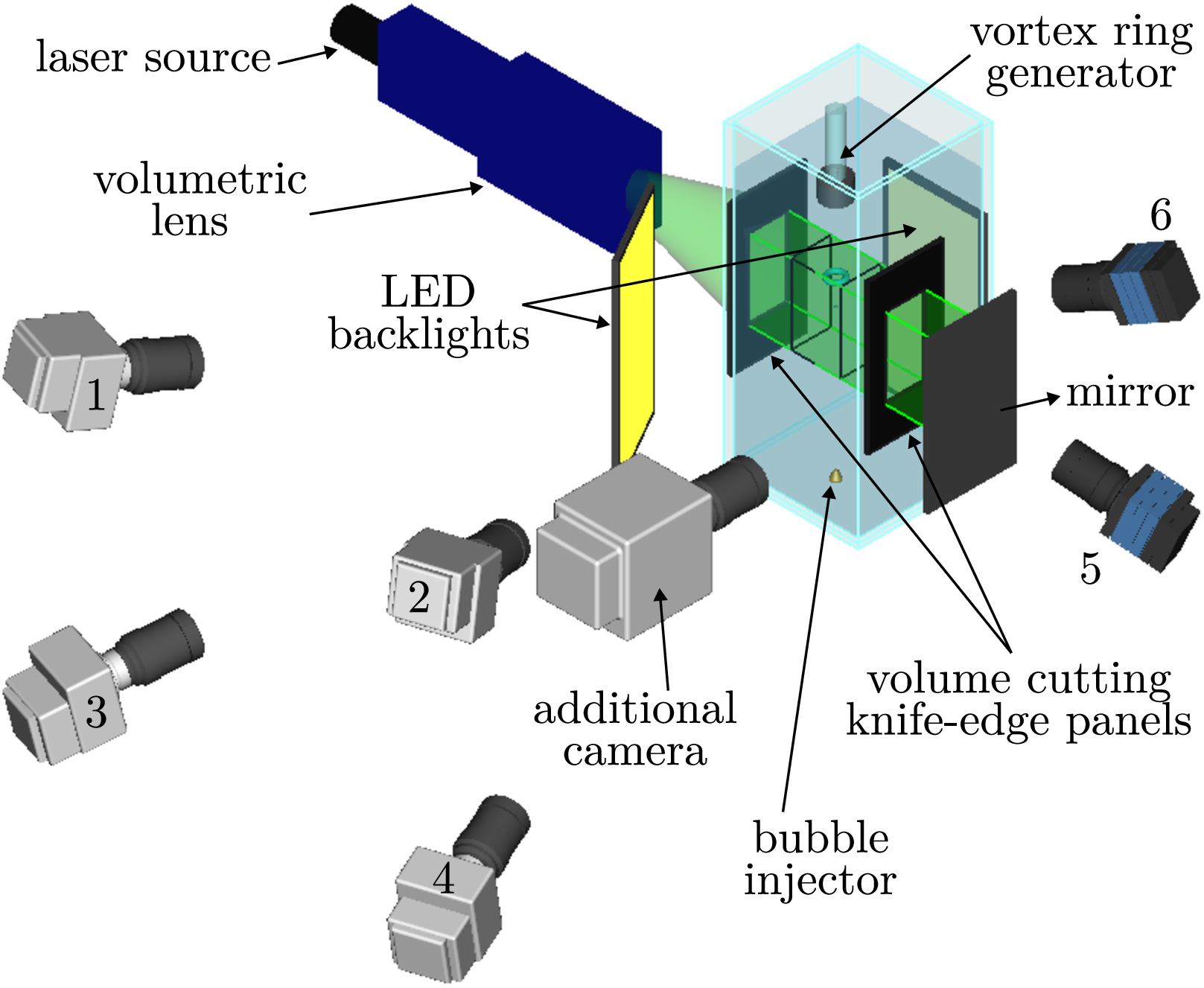}
        \caption{}
    \end{subfigure}    
    \hspace{1.5em}
    \begin{subfigure}{.5\textwidth}
        \includegraphics[width=\linewidth]{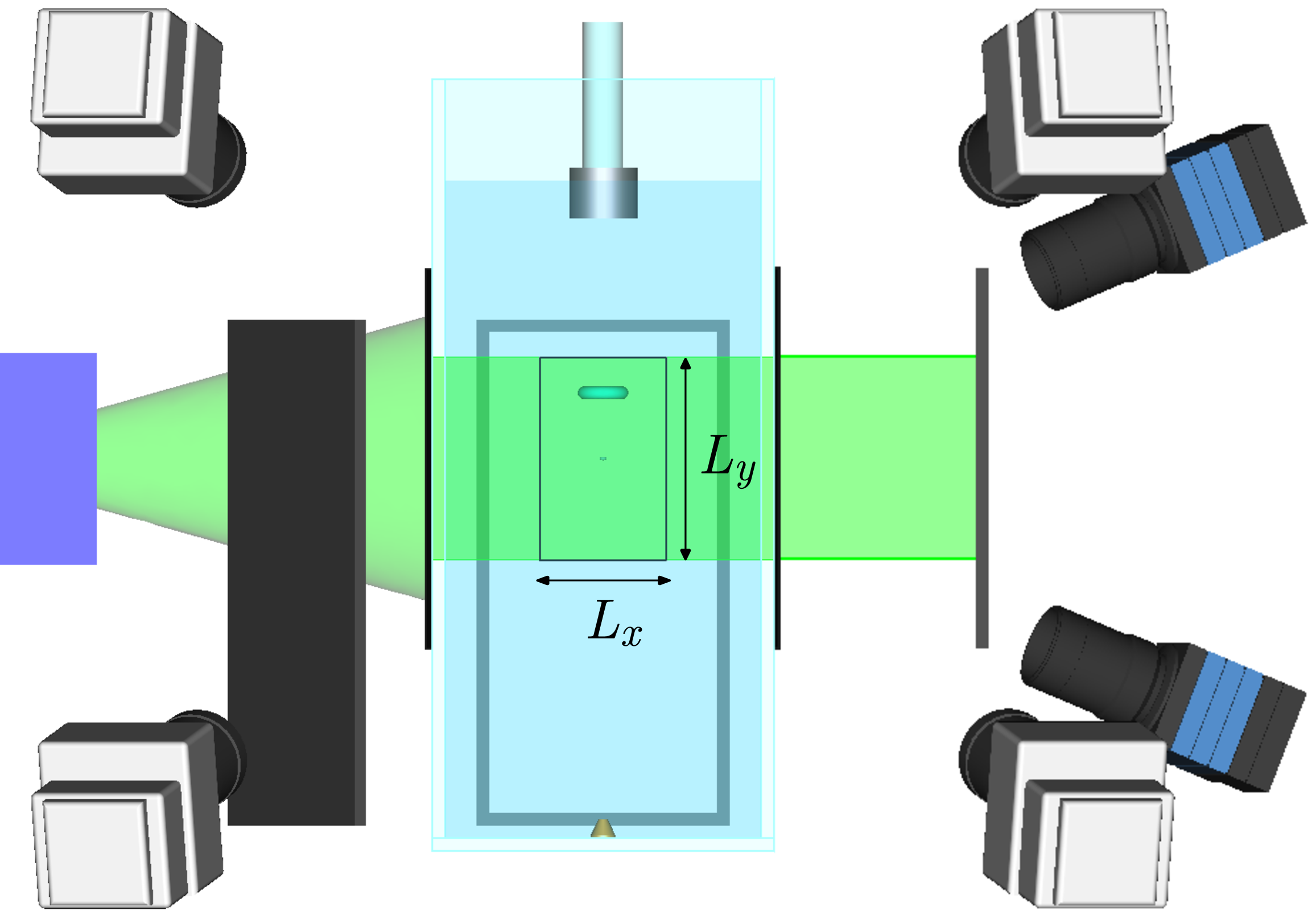}
        \caption{}
    \end{subfigure}
    \caption{(a) Isometric and (b) frontal view of experimental facility. Cameras used to record the volume of interest are numbered 1 to 6 in (a).}
    \label{fig:facility}
\end{figure}

The experiments were carried out at the Institute of Fluid Mechanics of Toulouse (IMFT). The experimental facility consisted of a $0.25\times0.25\times0.6$ m$^3$ tank filled with distilled water, seeded with 56 $\mu$m polyamide particles covered with rhodamine 6G (see figure~\ref{fig:facility}).  Tank temperature was maintained equal to 20$^{\circ}$C in order to keep constant water density, viscosity and refractive index. The vortex ring was generated by injecting a transient water jet from a pressurised reservoir into the still water tank. The jet orifice was located at the end of a 200 mm long, 30 mm inner diameter rigid tube and submerged well below the water free surface to ensure a well-controlled formation process~\citep{gharib1998universal}. It was aligned with the air inlet so that the collision of the bubble and the vortex ring was forced head-on. The jet, also seeded with fluorescent particles, was controlled by means of a solenoid valve, whose opening and closure times could be remotely adjusted. The circulation of the vortex ring was controlled using a 10 mm diameter orifice, with injection pressures between 1.5 and 3 bar and valve opening times ranging from 30 to 100 ms, yielding Reynolds numbers of $8\times10^3\lesssim Re \lesssim 15\times10^3$ and core-to-vortex radii ratios $a_0/R_0 \approx 0.3$ to $0.4$.

The bubble was released from the bottom using a syringe, a conical nozzle and a very low flow rate.  The nozzle diameter was adjusted depending on the desired bubble volume to guarantee the expected vortex-to-bubble size ratio, $R_0/R_b\simeq 4.5$. The air nozzle diameter was 5.5 mm, yielding bubble radii $R_b$ of $3.65 \pm 0.05$ mm. This parameter combination led to $R_0/R_b\approx4.5$, providing $0.4\lesssim We\lesssim1$, $1.6\lesssim Bo\lesssim 1.8$, and $615\lesssim Ga\lesssim690$. Upon release, the bubble would activate a photodiode sensor whose signal would then trigger the recording, as well as open the valve after a user-preset delay. Thanks to the repeatability of the experiment, the delay was adjusted so that most of the interaction between the bubble and the vortex ring happened within the field of view of all cameras.

The experiments were recorded at 1496 Hz acquisition frequency, using six high speed cameras: four VEO640L Phantom cameras from Vision Research$\textsuperscript{\textregistered}$ (2560$\times$1600 pixels) equipped with 200 mm F/16 Nikon lenses placed on a bench 100 cm away from the frontal face of the tank (cameras 1-4 in figure~\ref{fig:facility}a, shown also in the frontal view in figure~\ref{fig:facility}b); and two PCO DIMAX HS4 (2016 $\times$ 2016 pixels windowed to 1632 $\times$ 2016 pixels) equipped with 100 mm F/16 Nikon lenses located at one rear corner of the tank for a lateral view (cameras 5 and 6 in figure~\ref{fig:facility}a). Details on how the images obtained from these 6 cameras were calibrated may be found in Appendix \ref{app:calib}. To avoid reflections from the laser beam on the complex bubble interface and highlight the light scattering in the fluorescent seeding particles, the camera lenses were equipped with optical high-pass filters (540 nm). The camera setup provided multiple points of view, which optimised the tomographic measurements. A two-angle manual Scheimpflug mount was used between the lens and the sensor of each camera to compensate for the angular positioning required by the setup, ensuring minimal image distortion. In this manner, the horizontal angles between the optical axes of the frontal cameras 1-2 and 3-4 were $40^\circ \pm 1^\circ$. The vertical angles of the cameras 1-3, 2-4 and 5-6 were also $40^\circ \pm 1^\circ$, and the horizontal angles between the optical axes of the frontal and side cameras 2-6 and 4-5 were $97^\circ \pm 1^\circ$.
Two LED backlight sources were also included to obtain the bubble deformations and the eventual bubble breakup process via shadowgraphy. The illumination for the 4D-LPT images was provided by a 527 nm, 60 mJ/pulse double-pulsed laser (Photonics$\textsuperscript{\textregistered}$, ref. DM60-527-DH), expanded and collimated with a volume optics module from LaVision GmbH. Combined with a set of knife-edge panels, it illuminates only fluorescent tracer particles present in the $ 100\times 160\times 80$ mm$^3$ ($ L_x\times L_y\times L_z$) volume of interest (figure~\ref{fig:facility}b), optimizing measurements resolution with respect to the particle-per-pixel (ppp) limitations.
Moreover, a first surface mirror was placed in front of the laser on the other side of the tank to increase the fluorescence light of the particles in the volume of interest and to illuminate all the bubble surroundings, avoiding shadow effect. The measurement volume was located at mid-height of the tank to ensure that, when reaching it, the vortex ring was already properly formed and the bubble had reached its quasi-terminal rise path and velocity. However, it was placed closer to the front wall of the tank ($\sim40$ mm distance) to minimise turbidity caused by the particles present between the measuring volume and the wall, thus optimising optical access for the 4 front cameras, yet at a distance sufficient to avoid wall effects ($\sim4R_0\approx18R_b$). An additional Vision Research$\textsuperscript{\textregistered}$ V2012 high-speed camera (see figure~\ref{fig:facility}a) was placed near the injection point of the bubble to record the bubble formation and pinch-off process. This allowed us to corroborate that the formation was quasi-static and to have a precise estimation of the bubble volume from the frame taken right after release, when the bubble was still axisymmetric. The images from this camera were calibrated using a 1 mm precision scale. See Appendix \ref{app:calib} for more information.

\subsection{Flow variables determination}\label{subsec:flow_variables}
After characterising the position and size of the vortex, we may calculate the different flow variables involved in the problem. See Appendices \ref{app:STB} -\ref{app:processing} for technical details. In particular, the circulation of the vortex ring was computed by performing a surface integral of the vorticity in an enlarged area enclosing each core (left/right) of the vortex. Each area was defined as a vertical plane extending the full vertical length of the grid with a radial length covering the entire domain in the positive radial direction from $r_v$ (i.e., $r_l$ or $r_r$, respectively). Therefore, these areas constitute half-planes which are defined at each value of $\theta$. As verification, we also computed the circulation using a line integral of the axial velocity around the boundary of the above-mentioned area. Given the Stokes theorem, both methods are equivalent \citep{batchelor2000introduction}, and they gave the same results in our experiments when computed as:
\begin{equation}
\Gamma(t,\theta)=\int_\Sigma \vec{\omega}.\vec{n} \, d\sigma=\oint \vec{u} . d\vec{l},
\end{equation}
where $\vec{u}$ and $\vec{\omega}$ are the velocity and vorticity of the flow, $\vec{n}$ the  vector normal to the plane of surface $\Sigma$, bounded by $d\vec{l}$. 

The vortex ring may be further analysed by calculating the kinetic energy and the enstrophy of the flow. These three-di\-men\-sional (i.e., volumetric) scalar fields are defined as follows:
\begin{equation}
\begin{gathered}
    E_k(t)= \frac{1}{2}\int_V\rho |\vec{u}|^2dV,\\
    \Omega(t) = \frac{1}{2}\int_V|\vec{\omega}|^2dV=\frac{1}{2}\int_V(\omega_r^2+\omega_\theta^2+\omega_y^2)dV=\Omega_r+\Omega_\theta+\Omega_y, \\
    \label{eq:eqs_integral}
\end{gathered}
\end{equation}
where $V$ is the volume of liquid, and $\Omega_r,~\Omega_\theta,~\Omega_y$ are the contributions to the enstrophy of the radial, azimuthal, and vertical components of vorticity ($\omega_r,~\omega_\theta,~\omega_y$), respectively. We may directly adapt these three-dimensional definitions to our cartesian-grid-like domain with
\begin{equation}
    E_{k,V}(t)= \frac{1}{2}\rho\sum_{i=1}^{n_V}|\vec{u}|_i^2 \, V_i,
    \hspace{3em}
    \Omega_{j,V}(t) = \frac{1}{2}\sum_{i=1}^{n_V}(\omega_j)^2_i \, V_i, \label{eq:eqs3D}
\end{equation}    
where $\vec{u}_i$ and $\vec{\omega}_i$ indicate the velocity and vorticity at voxel $i$, respectively, $n_V$ is the number of cells in the part of the grid considered as the integration box, $V_i$ is the volume of the cell $i$ in the grid, and $j$ represents each contribution of the vorticity in the three-di\-men\-sional field (radial, azimuthal or axial). We did not take into account the whole grid, but only a reduced volume (highlighted in figure~\ref{fig:block_diagram}b), to avoid adding noise to the results. To select the optimal grid size, we performed a sensitivity analysis of its effect on the energy and enstrophy calculation. Given the cylindrical symmetry of the problem, we calculated both 3D variables in several cylindrical volumetric domains of growing radius around the vortex centre. We selected the cells inside a cylinder around the vortex whose radius was as small as possible, but sufficiently large to avoid critical loss of information. Note that only the cells corresponding to the liquid were taken into account in this calculation, masking out the cells covered by the 3D-reconstructed bubble.\\

To be able to compare our results with the 2D studies available in the literature, we may also compute the planar kinetic energy and the planar (azimuthal) enstrophy contained within every angular division of the flow field. This will also be of interest to analyse the azimuthal region near the bubble, $\theta_{bubble}$ (see figure~\ref{fig:problem_sketch}a), separately from the unperturbed zones of the vortex, thus isolating its impact more clearly. Furthermore, since the bubbles move before, during and after the interaction with the vortex ring, this approach will provide all the flow information around the bubble region, regardless of the potential bubble azimuthal displacement. Note that depending on the information we want to extract, we may need to analyse only half a plane ($\theta$, to see the local effect of the bubble at the core) or gather the results from a complete azimuthal cut of the grid ($\theta$ and $\theta+180^\circ$, e.g. to compare with typical 2D-PIV results). For this reason, although we sweep the grid for $\theta$ varying between 0 and 180$^\circ$ (full plane), we collect the results separately for each half-plane ($\theta$ between 0 and 360$^\circ$), being our planar equations defined on a 2D domain that covers all values of $y$, but only $r$ contained within the area corresponding to a vertical cut of the equivalent volumetric domain used for the 3D calculations in the same experiment. Thus, we define the planar counterparts of equation~\eqref{eq:eqs3D} as

\begin{equation}
    E_{k,A}(t,\theta)= \frac{1}{2}\rho\sum_{i=1}^{n_A}|\vec{u}_{ry}|_i^2\Delta r \Delta y,
    \hspace{3em}
    \Omega_{j,A}(t,\theta) = \frac{1}{2}\sum_{i=1}^{n_A}(\omega_j)^2_i\Delta r \Delta y, \label{eq:eqs2D}\\
\end{equation} 
where $n_A$ represents the number of cells in the reduced area of the plane considered, $|\vec{u}_{ry}|_i=\sqrt{u_{r,i}^2+u_{y,i}^2}$ and $(\omega_{j})_i$ are the in-plane velocity of the flow and its $j$-component of the vorticity at cell $i$ in the grid, respectively, and $\Delta r$ and $\Delta y$ are the grid cell size in the radial and vertical directions. Notice that we have included the three components in equation~\eqref{eq:eqs2D}, $\omega_j$ with $j=r, \theta, y$, since all of them can be extracted in a plane from the 3D measurements.

To directly compare the values of the volumetric flow variables and their planar counterparts, we averaged them over such an area or volume. Thus, the kinetic energy and contribution of the -$j$- component of the vorticity to the enstrophy, averaged in a plane of area $A=n_A\Delta r\Delta y$ with orientation $\theta$ at time $t$, referred to as specific kinetic energy and enstrophy, respectively, write
\begin{equation}
    E^{s}_{k,A}(t,\theta)= 
    \frac{1}{n_A\Delta r\Delta y}E_{k,A}(t,\theta),
    \hspace{1em}
    \Omega^{s}_{j,A}(t,\theta)=
    \frac{1}{n_A\Delta r\Delta y}\Omega_{j,A}(t,\theta).\\
    \label{eq:eqs_averaged2D}
\end{equation}
Similarly, the specific kinetic energy and the contribution of the $j$-component of the vorticity to the enstrophy in a volume $V=n_VV_i$ at time $t$ can be calculated as
\begin{equation}
    E^{s}_{k,V}(t)= \frac{1}{n_VV_{i}}E_{k,V}(t),
    \hspace{3em}
    \Omega^{s}_{j,V}(t) = \frac{1}{n_VV_i}\Omega_{j,V}(t),
    \label{eq:eqs_averaged3D}\\
\end{equation}
with $E_{k,V}(t)$ and $\Omega_{j,V}(t)$ defined in equations~\eqref{eq:eqs3D}, being all cells of equal size. Note that for the calculation of the planar kinetic energy and enstrophy, to be coherent with their 3D counterparts, a reduced area was used instead of the full planar grid, equivalent to a planar cut of the cylindrical subvolume used in the volumetric case (see figure~\ref{fig:block_diagram}b). The decision of considering only the central region of the grid, where most of the vorticity and velocity are concentrated, for the surface/volumetric integration of these variables is essential to avoid underestimating the averaged values.

\FloatBarrier
\section{Results}\label{sec:results}
\begin{figure}
    \centering
\includegraphics[width=0.85\linewidth]{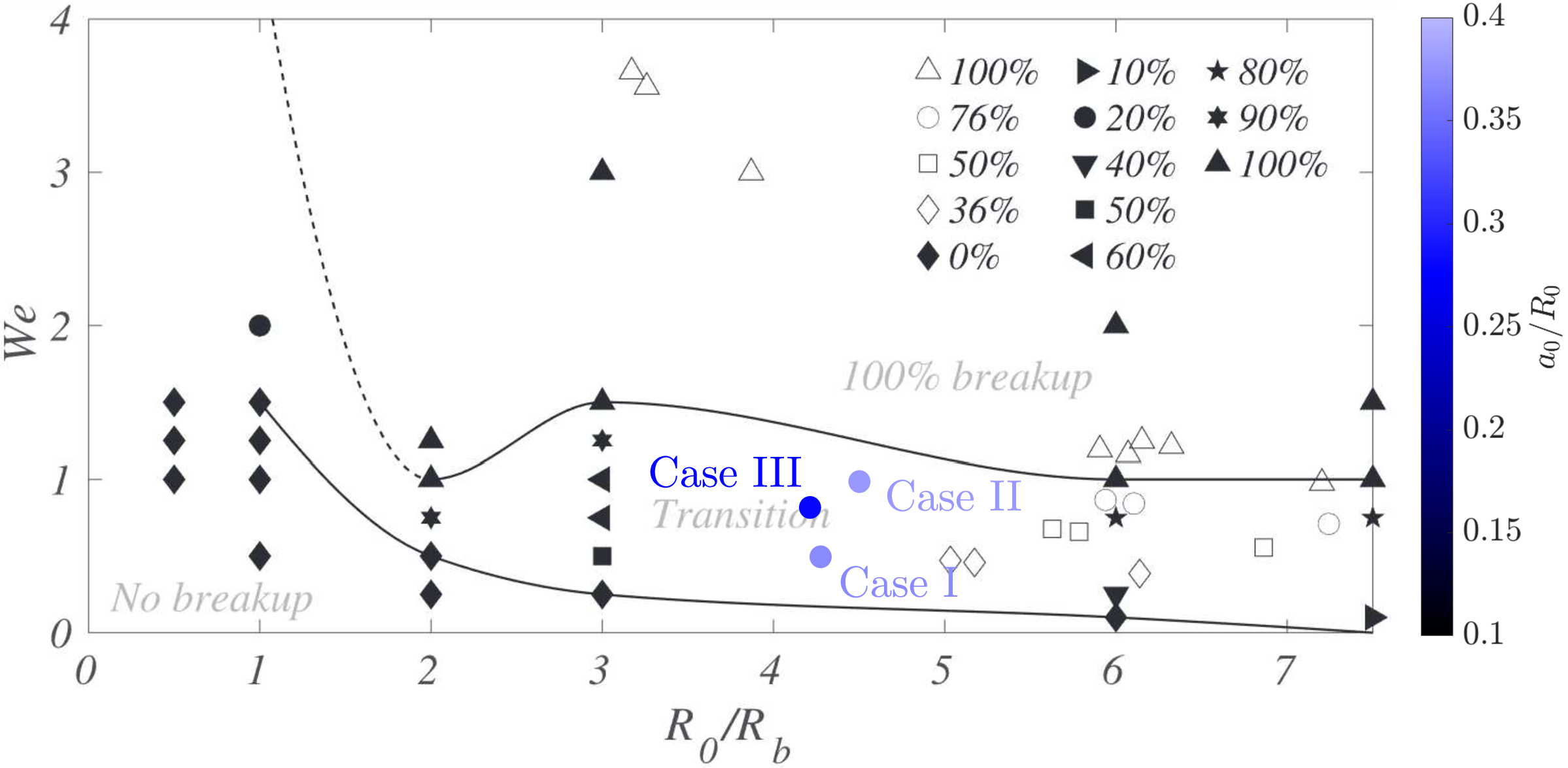}
    \caption{Map of the regions where bubble breakage is observed in the $R_0/R_b$\textit{\textendash}$We$ plane \citep{foronda2021deformation}. Experimental data are represented by hollow symbols indicating the bubble breakup probability. Solid symbols indicate the numerical simulations, with each symbol consisting of at least ten different simulations varying the initial position of the bubble with respect to the vortex. The solid lines separate three regions where bubbles break in all the cases, in none or only in a fraction of them. The three interaction cases studied here were included in the map, with all data coloured by the vortex thickness ratio $a_0/R_0$.}
    \label{fig:Foronda}
\end{figure} 
The present study focuses on the three representative cases corresponding to distinct bubble-vortex interaction regimes (weak interaction, capture--transport--release, and bubble breakup), see Fig.~\ref{fig:Foronda}. These cases were selected as the most reproducible and illustrative behaviours, after performing numerous preliminary experiments varying the vortex orifice diameter, the injection pressure, and the valve aperture time, as well as the bubble size. As a means of validation of our vortex identification algorithm, we completed several measurements with a vortex ring only, with no bubble interaction.

\subsection{Description of the interactions explored}\label{sec:Intro_cases}
To evaluate the three-di\-men\-sional experimental approach, we focus on three specific bubble-vortex interaction cases: Case I, in which a weak interaction between the bubble and the vortex ring takes place; Case II where the bubble is captured and dragged by the vortex; and Case III with the bubble breaking up. The three cases studied are included in figure~\ref{fig:Foronda}, which reproduces the transition map from \cite{foronda2021deformation}. In that study, numerical simulations were performed on bubble-vortex interactions at a fixed $Re = 15000$ and a core ratio of $a_0/R_0 = 0.1$ (thin vortex rings), while experiments were performed at $Re \in (9000, 20000)$. The figure illustrates the breakup efficiency of experiments (hollow symbols) and numerical simulations (solid symbols), and transition curves are obtained depending on breakup probability. In this work, we intend to solve the three-dimensional flow field in three different characteristic interaction cases while keeping $R_0/R_b$ approximately constant. Thus, we first considered a low-Weber-number case, in which the bubble was neither trapped nor broken by the vortex. In the second case, we increased the Weber number while attempting to keep $a_0/R_0$ equal to that of Case I, in order to achieve vortex trapping of the bubble without breakup. To highlight the effect of the vortex core ratio, we aimed to induce bubble breakup at $We\sim 1$, by reducing $a_0/R_0$. Thus, we observed bubble trapping and breakup in Case III. It should be noted that the three cases under study here are located in the transition region, where eventual bubble breakup is not a deterministic phenomenon. In this way, two experiments with $We\sim 1$ and identical values of $R_0/R_b$ and $a_0/R_0$ do not necessarily lead to the same outcome (bubble breakup or trapping). Therefore, rather than discussing the influence of parameters on the observed dynamics, we take an experimental approach to characterise the unsteady and three-dimensional features of the three different types of interaction selected.
\begin{figure}
    \centering
    \includegraphics[width=\linewidth]{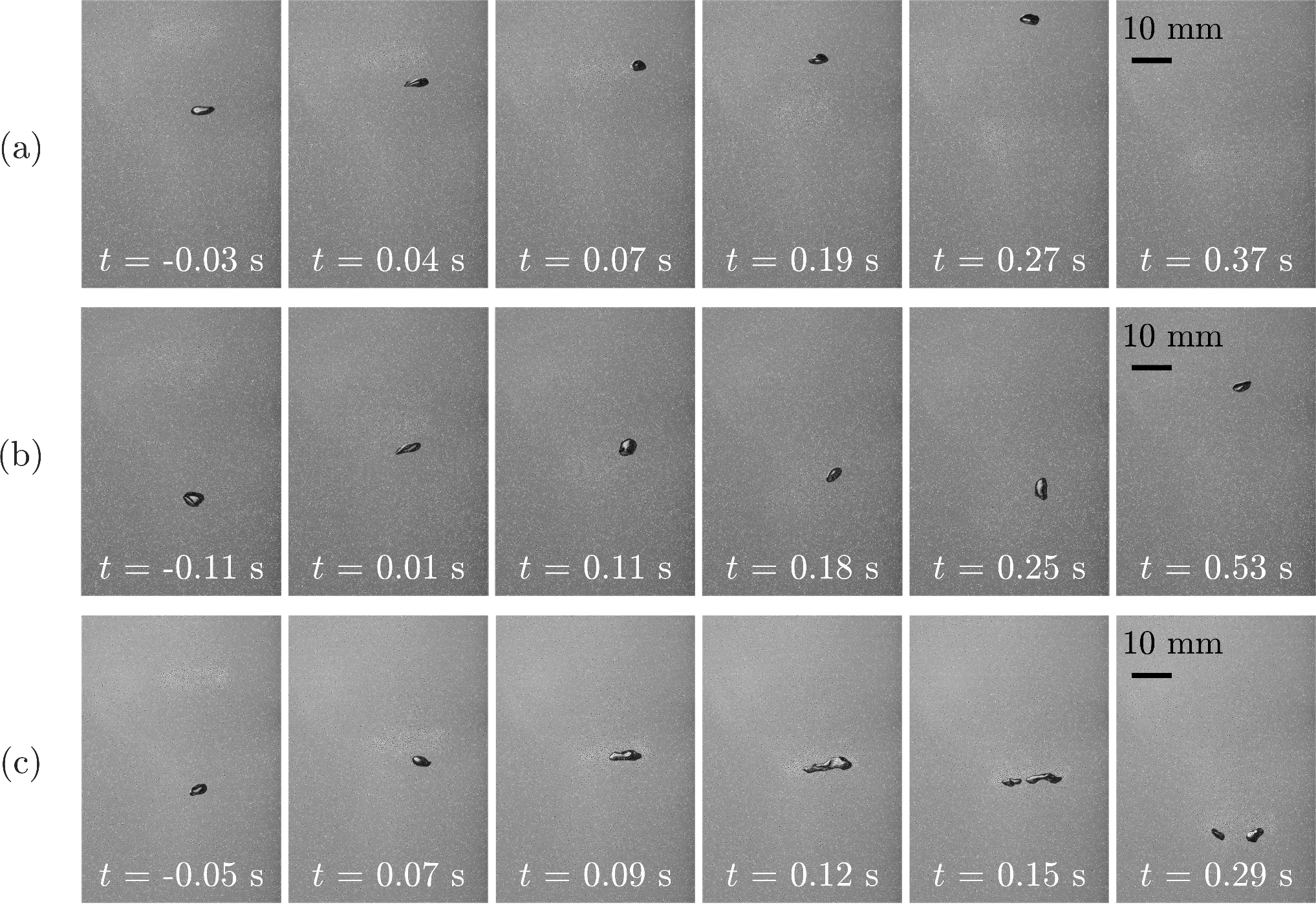}
    \caption{Time sequence of the front view images (camera 1) showing representative cases of three different types of interaction covered in this work. (a) Case I: Weak interaction; (b) Case II: Capture, drag and release of the bubble by the vortex ring; (c) Case III: Binary bubble breakup. Supplementary movies related to this figure (Movie2\_raw\_caseI.mp4, Movie3\_raw\_caseII.mp4, and Movie4\_raw\_caseIII.mp4) are accessible at journals.cambridge.org.}
    \label{fig:raw_experiments}
\end{figure}
\begin{table}
    \centering
    \begin{tabular}{cccccccccc}
        Case & Description & $R_0/R_b$ & $a_0/R_0$ & $a_0/R_b$ & $Re$ & $We$ & $Bo$ & $Ga$ & $Fr$\\
        \hline
        I & Weak interaction & 4.3 & 0.37 & 1.59 & 8900 & 0.41 & 1.57 & 618.5 & 1.35\\
        II & Capture, drag and release & 4.5 & 0.38 & 1.71 & 14600 & 1.00 & 1.81 & 690.5 & 1.25\\
        III & Bubble breakup & 4.2 & 0.28 & 1.18 & 11800 & 0.77 & 1.75 & 671.5 & 1.28\\
    \end{tabular}
    \caption{Dimensionless parameters describing the different types of interactions studied. Based on the sub-pixel accuracy of the measurement technique (described in Appendix 
    ~\ref{app:STB}), the errors in the flow-related parameters and the bubble size are below $3\%$ and $1\%$, respectively.}
    \label{tab:cases}
\end{table}
Figure~\ref{fig:raw_experiments} depicts raw images captured at various time instants for each case. Both the bubble and the vortex ring can be observed in the figure: the bubble rises and the vortex ring descends until they interact. The type of interaction results from the values of the governing parameters, listed for each case in Table~\ref{tab:cases}. Additionally, the initial characteristics of the vortex rings are detailed in Table~\ref{tab:initial}. Note that the uncertainties in these parameters, and in the rest of the results throughout the manuscript, were estimated using standard propagation methods based on the literature \citep{ho1983precision, stanislas2008main, kahler2016main, schanz2016shake, sciacchitano2016piv}. The error propagation analysis ensures that the cumulative relative uncertainty in velocity ranges between 2 and 10\% in the region near or far from the vortex, respectively.

Figure \ref{fig:raw_experiments}(a) shows a weak bubble-vortex interaction (Case I). As both the bubble and vortex ring come into proximity, the bubble trajectory is slightly diverted by the flow induced by the vortex. Since $We$ is well below unity, surface tension forces dominate over the inertial forces induced by the vortex. In configurations where the vortex ring is relatively larger than the bubble ($R_0/R_b=4.3$), interaction events at such low Weber numbers are typically characterised as ``Free Crossing'' events, where the bubble is not successfully trapped or broken by the vortex core~\citep{foronda2021deformation}. Thus, the flow cannot produce a sustained stretching strong enough to deform nor break the bubble. Under these conditions, the bubble undergoes only moderate, reversible deformation. Moreover, the relatively thick vortex core ($a_0/R_0 = 0.37$) presents a smoother velocity distribution than thinner vortex rings. This further reduces the ability of the vortex to hold the bubble within its core. Thus, the bubble is accelerated towards the vortex core and reduces its rise velocity, but is not captured. After this, the bubble continues its rise upwards, restoring its path and terminal velocity after a short transient.
\begin{table}
    \centering
    \begin{tabular}{cccccc}
        Case & $R_0$ (mm) & $a_0$ (mm) & $\Gamma_0$ (cm$^2$/s) & $\Omega_{V,~0}^s$ (s$^{-2}$) & 
$E_{k,V,~0}^s$ (J/m$^3$) \\
        \hline
        I & 15.9 & 5.89 & 94.1 & 222.8 & 2.68  \\
        II & 16.6 & 6.32 & 145.6 & 326.7 & 5.60  \\
        III & 15.2 & 4.34 & 118.6 & 533.4 & 5.87 \\
    \end{tabular}
    \caption{Characteristic values of the vortex ring variables at $t=t_0$ for each case.}
    \label{tab:initial}
\end{table}
In Case II, the interaction is more intense, as illustrated in figure~\ref{fig:raw_experiments}(b), because both $We$ and $Re$ numbers are larger than in the previous case, while the values of the rest of the parameters remain practically unchanged. On one hand, a higher $Re$ indicates that the ring is energetic and coherent, able to advect the bubble once it is captured. On the other hand, a higher $We$ promotes bubble deformation. In particular, in this case, $We=1$, meaning that the inertial stresses of the vortex ring are comparable to surface tension. This leads to bubble deformation, but not necessarily to its fragmentation. Therefore, the bubble can be seen to be drawn towards and captured inside the core region, it remains moving downwards within the ring for a certain time interval until it is finally released and continues its rise.

Case III, illustrated in figure~\ref{fig:raw_experiments}(c), corresponds to the most intense interaction: the bubble is trapped in the vortex core and remains inside for a relatively long time. In this case, inertial stresses are sufficient to overcome the restoring action of surface tension ($We\simeq0.8$), so the bubble elongates azimuthally along the ring until it eventually collapses into two fragments that stretch along the core, as will be demonstrated later. Note that, although $We$ is lower than in Case II (but still close to unity), it can lead to bubble fragmentation. The fact that the core is thinner ($a_0/R_0=0.28$ and $a_0/R_b=1.18$ in Case III, while $a_0/R_b=1.71$ in Case II) is also a key feature promoting bubble breakup, since the ring’s straining region is more concentrated, producing sharper, more powerful localised stretching of the bubble due to more abrupt velocity gradients. 

\begin{figure}
    \centering
    \includegraphics[width=\linewidth]{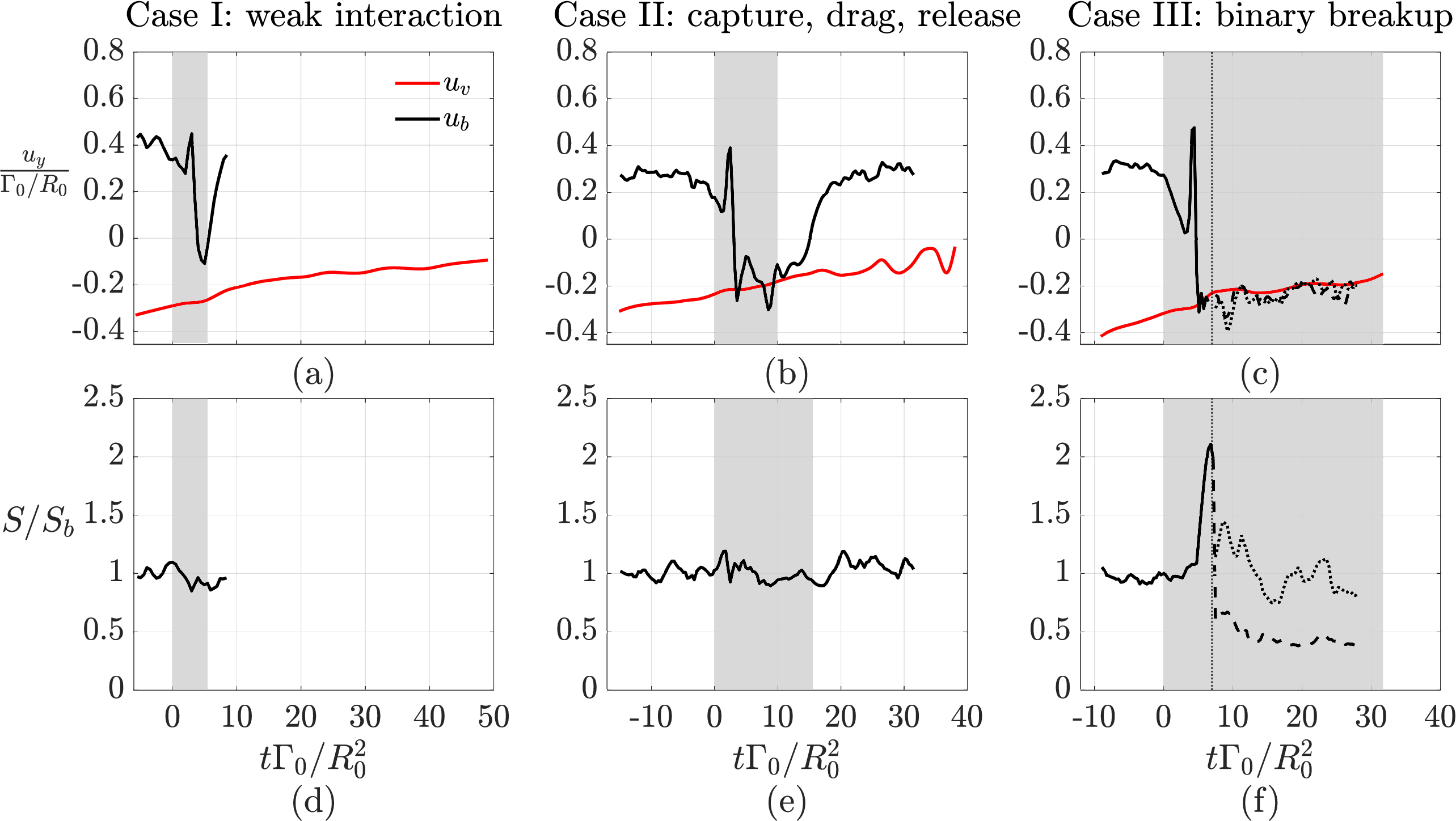}
    \caption{Top row: vertical velocity, $u_y$, of the vortex ring ($u_v$, red) and the bubble ($u_b$, black) for the cases regarding (a) weak interaction, (b) capture, drag and release, and (c) bubble breakup, respectively. Bottom row: temporal evolution of the bubble surface ratio, $S/S_b$, for the cases regarding (d) weak interaction, (e) capture, drag and release, and (f) bubble breakup, respectively. Dotted and dashed lines in (c) and (f) show daughter bubbles' velocities and surfaces, respectively. The interaction time is highlighted in grey.}
    \label{fig:compare_kinematics}
\end{figure}

As a first stage of the analysis of three cases, we present in figure~\ref{fig:compare_kinematics} the evolution of the vertical velocities, $u_y(t)$, of both the vortex ring, $u_v(t)$, and the bubble, $u_b(t)$ (see figure~\ref{fig:compare_kinematics}a-c), in dimensionless form, as well as the bubble surface ratio, $S(t)/S_b$ (see figure~\ref{fig:compare_kinematics}d-f), with $S_b$ the initial bubble surface obtained from the bubble formation recordings of the additional camera in figure~\ref{fig:facility}(a), and $S(t)$ the surface of the reconstructed bubble at time $t$. In all cases, the vortex slows down as it moves downstream. The bubble accelerates quickly and suddenly, and then its velocity decreases sharply when it encounters the vortex ring. Note that the availability of time-resolved data for the trajectory of the bubble allows us to calculate the duration of the interaction: we define the beginning of the interaction, $t=0$, from the moment the bubble deviates from its rising path due to the presence of the vortex ring, and its final, $t_f$, when the bubble turns back to its stationary rise and leaves its influence, being the interaction time $t_i=t_f$. We marked the starting and final points of the interaction time (shaded area in figure~\ref{fig:compare_kinematics}) by looking for peaks and inflection points in the bubble velocity, representing strong, fast deviations from the previous path and thus indicating the influence of the vortex ring on the bubble. 

The bubble recovers its terminal velocity after being released by the vortex in Cases I and II (figures~\ref{fig:compare_kinematics}a, b). In Case III (figure~\ref{fig:compare_kinematics}c), the bubble is trapped by the vortex core, which dramatically slows its velocity and causes it to break into two smaller bubbles (dashed and dotted lines in figures~\ref{fig:compare_kinematics}c, f). These daughter bubbles are then dragged by the vortex at its own velocity. Regarding the bubble surface, in Cases I and II (figures~\ref{fig:compare_kinematics}d, e), it remains nearly constant because the bubble is barely deformed during its interaction with the vortex, as will be described later. Nevertheless, in Case III (figure~\ref{fig:compare_kinematics}f), the mother bubble is strongly elongated within the vortex core before breaking up ($S/S_b$ increases up to $\approx2$). The bubble splits into two unequal daughter bubbles: one with approximately the same area as the mother bubble (dotted line), and another with nearly half the area (dashed line). The surfaces of both bubbles decrease slightly over time, suggesting that their surfaces are being restored, becoming more spherical, due to surface tension effects, as these bubbles are smaller than the mother bubble (and thus their corresponding Weber numbers are also smaller).

In the following, we will analyse in detail the kinematics and dynamics of each type of interaction obtained from the experimental data. While the overall outcomes of these interactions are consistent with the transition map proposed by \cite{foronda2021deformation}, thus supporting the validity of the present experimental conditions, the time-resolved three-dimensional measurements provide access to additional features of the interaction that cannot be captured using planar approaches. In particular, beyond the global classification of the interaction regimes, the present results allow us to resolve the spatial and temporal distribution of key quantities such as bubble deformation and enstrophy. For Case I, the interaction is shown to be highly localised in the azimuthal direction, with transient variations confined to the bubble plane. For Case II, the measurements reveal the progressive transition from a localised perturbation to a global deformation of the vortex ring during the capture and release process. For Case III, the three-dimensional reconstruction enables the quantification of the azimuthal elongation prior to breakup and provides insight into the redistribution of enstrophy, including the increase of radial and axial vorticity components associated with the formation of smaller-scale structures.

\subsection{Case I: weak interaction}\label{subsec:caseI}
\begin{figure}
    \centering
    \includegraphics[width=\linewidth]{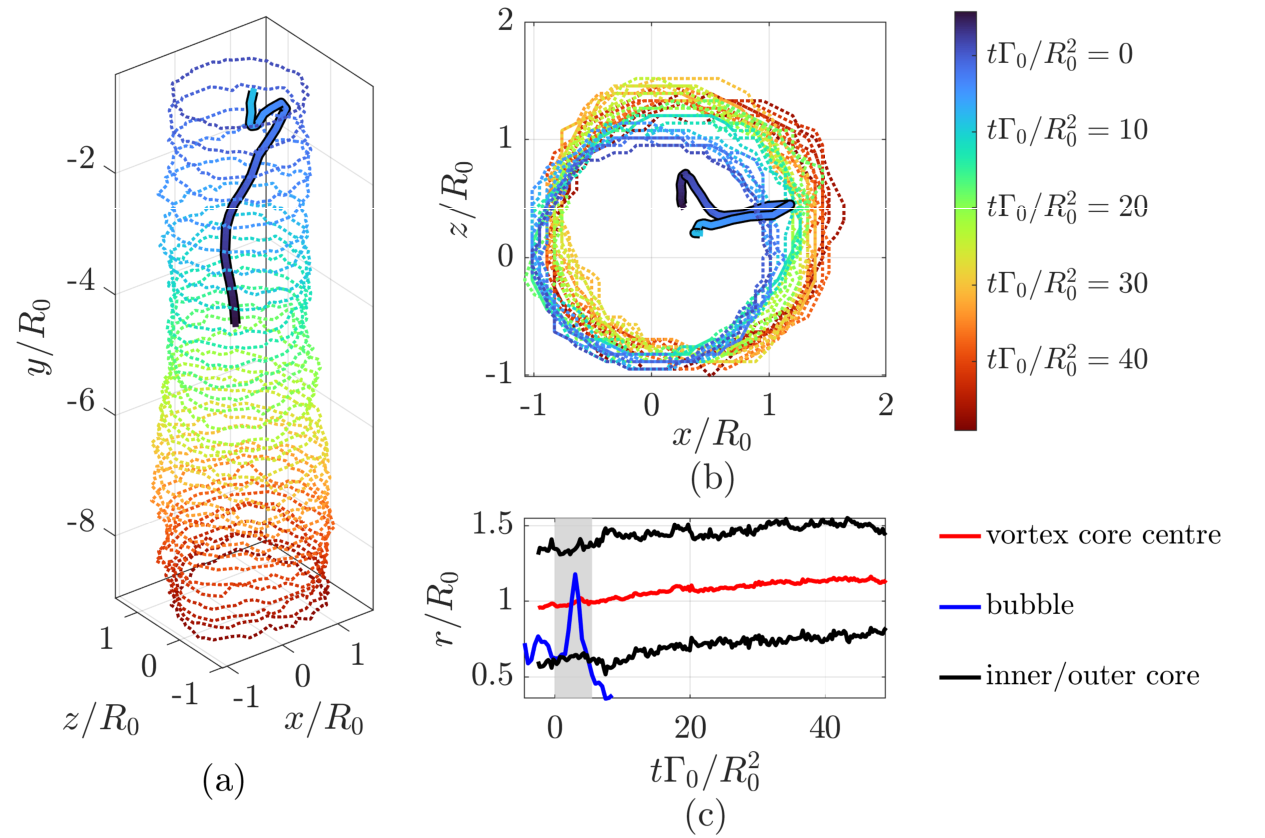}
    \caption{(a) Three-dimensional and (b) top views of the bubble trajectory (thick solid line) and time evolution of the shape and position of the vortex ring core centre (thin dotted lines) in the weak interaction case (Case I). Both are coloured by time (see colourbar), with the bubble travelling upwards and the vortex downwards (counterflow). (c) Temporal evolution of the azimuthally averaged radial position of the vortex core centre and inner/outer limits, together with the bubble centroid. The shaded area in panel (c) corresponds to the duration of the interaction.}
    \label{fig:trajectory_WI}
\end{figure}
Figure \ref{fig:trajectory_WI} shows the temporal evolution of both the vortex ring core centre and the bubble position corresponding to a weak interaction. In particular, panels (a) and (b) illustrate the spatial evolution of both the vortex ring and the bubble, coloured by dimensionless time, $t\Gamma_0/R_0^2$. Note that we are able to solve the three-di\-men\-sional bubble trajectory as well as the position and the shape of the vortex ring: while the three-di\-men\-sional evolution is depicted in panel (a), the top view is shown in (b). The vortex ring core (thin dotted lines) propagates downwards, while the bubble (thick solid line) travels upwards. The bubble, initially rising vertically along the $y$-axis, is pulled laterally away from the axis towards the core of the approaching vortex ring. In particular, when both the bubble and the ring are close enough, the bubble quickly deviates to its core, but it escapes very soon. The interaction time is short according to figures~\ref{fig:trajectory_WI}(a), (b), where it is observed that the interval of time when the bubble and vortex coincide in both colour and position is brief. This interaction time is highlighted with a shaded area in figure~\ref{fig:trajectory_WI}(c), where the time evolution of the radial position of the bubble centroid (blue), the outer and inner limits of the vortex core (black), and its centre (red) are plotted. The bubble interacts with the vortex at $t\Gamma_0/R_0^2=0$, and quickly moves towards the core centre, i.e., the radial position increases. Subsequently, after a very short period, its radial position deeply decreases, and the bubble escapes the core at the interaction time, $t_i\Gamma_0/R_0^2\simeq6$. Regarding the vortex ring, it maintains an almost circular form at each moment (figure~\ref{fig:trajectory_WI}), with a radius $R(t)$ that slightly grows over time 
after the perturbation (red line in figure~\ref{fig:trajectory_WI}c). Moreover, the ring centre deviates towards the interaction position (see figure~\ref{fig:trajectory_WI}b). This minimal distortion is expected in a weak interaction scenario, which fails to induce significant and lasting azimuthal instability or fragmentation in the vortex core.
\begin{figure}
    \centering
    \includegraphics[width=\linewidth]{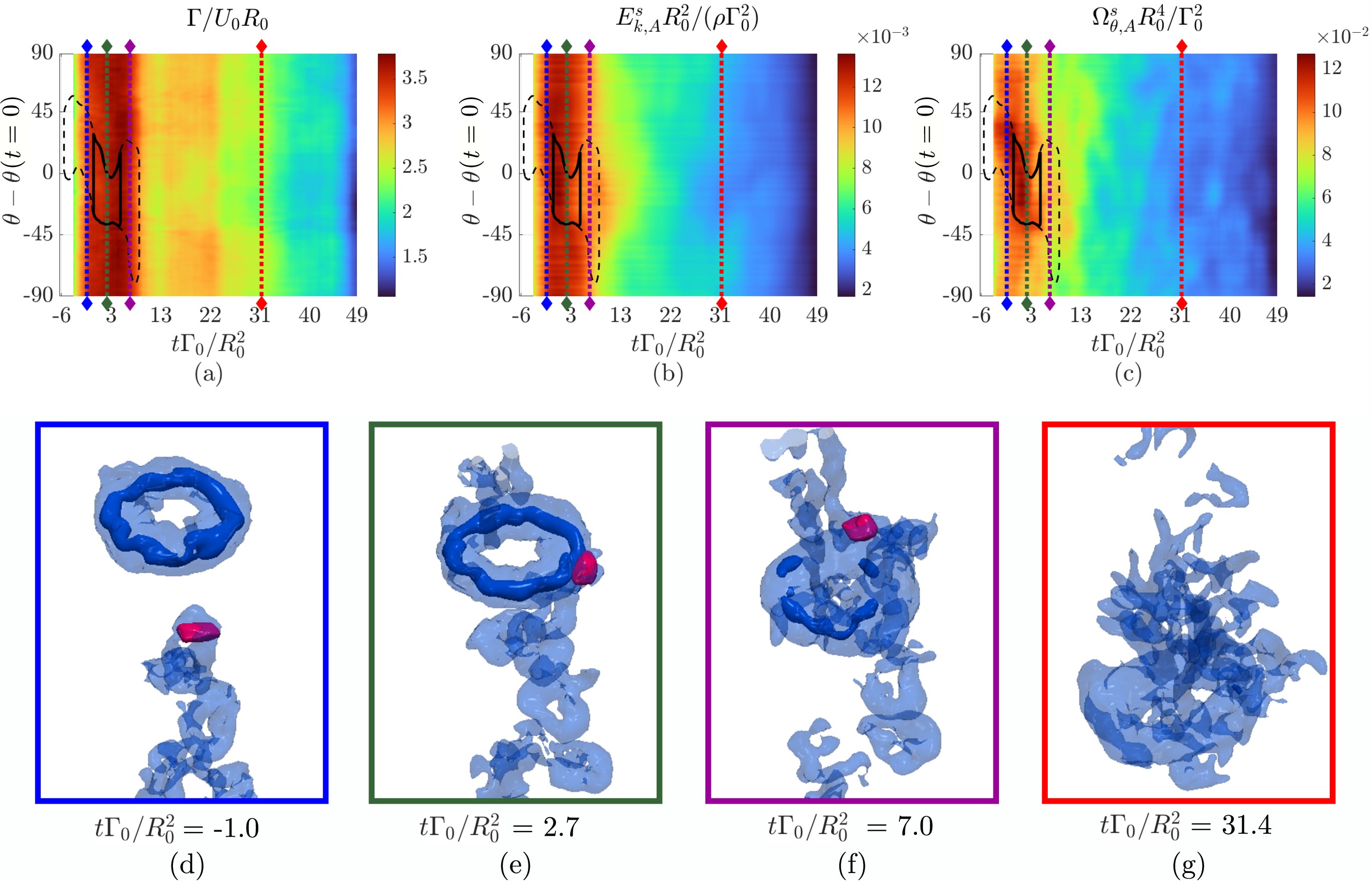}
    \caption{Temporal evolution of the azimuthal distribution of the (a) circulation, $\Gamma/(U_{0}R_{0})$, (b) planar, specific kinetic energy, $E_{k,A}^{s}R^{2}_{0}/(\rho \Gamma^{2}_{0})$, and (c) planar, specific azimuthal enstrophy, $\Omega_{\theta,A}^{s}R^{4}_{0}/\Gamma^{2}_{0}$, of the weak interaction case (Case I). The values correspond to planar azimuthal sections covering two cores with $\theta\in[0^\circ,180^\circ]$ (full plane), as it would be obtained in 2D-PIV measurements. The areas enclosed by black dashed lines mark the instants and azimuthal coordinates corresponding to the position of the bubble, $\theta_{bubble}$, while the solid ones correspond to its position while it remains inside the vortex core during the interaction. The dotted vertical lines indicate the instants corresponding to different snapshots throughout the experiment (d to g, matched by colour) for a better understanding of the effect of the interaction. The bubble surface is coloured in magenta, while blue represents the $\lambda_2$-criterion iso-surfaces (opaque blue: $\lambda_2 R^{4}_{0}/\Gamma^{2}_{0}=-1.08$; faded blue: $\lambda_2 R^{4}_{0}/\Gamma^{2}_{0}=-0.0072$). A supplementary movie related to this figure (Movie5\_caseI.mp4) is accessible at journals.cambridge.org.}
    \label{fig:map_WI}
\end{figure}

Figure \ref{fig:map_WI} presents dimensionless circulation, $\Gamma/(U_{0}R_{0})$ in (a), planar, specific kinetic energy,  $E_{k,A}^{s}R^{2}_{0}/(\rho \Gamma^{2}_{0})$ in (b) and planar, specific azimuthal enstrophy, $\Omega_{\theta,A}^{s}R^{4}_{0}/(\rho \Gamma^{2}_{0})$, in (c) as a $\theta$-time map. We also include three-dimensional snapshots corresponding to specific time instants (d-g). The angle $\Delta\theta=\theta-\theta(t=0)$ describes the position relative to the azimuthal coordinate of the centroid of the bubble at the beginning of the interaction. This means that $\Delta\theta$ defines the azimuthal extension of the bubble. Note that planar kinetic energy and azimuthal enstrophy in a particular $\theta$ and time correspond to equations~\eqref{eq:eqs_averaged2D} averaged in $(y,r)$ planes covering $180^\circ$ (full planes). The orientation of the planes where these magnitudes are calculated is varied from $-90^\circ<\Delta\theta<90^\circ$ to span the entire ring and locate the interaction at the centre of the domain. This planar analysis, covering the two cores of the vortex, is comparable to a classical 2D study.

It is important to observe that for any azimuthal location, after a rapid decrease following the interaction, all three magnitudes show a continuous, gradual decay over time due to the weakening of the vortex as it descends. This implies that the interaction is not highly disruptive to the overall strength and long-term coherence of the vortex ring, in agreement with previous studies~\citep{biswas2022interaction}. Moreover, it can be noted that, close to the bubble region, $\theta_{bubble}$, all magnitudes, but mostly enstrophy, are slightly perturbed along $\theta$, thus revealing the three-di\-men\-sional nature of the interaction. The sharp localisation indicates that the most intense modification occurs exactly where the bubble momentarily resides (black line), failing to spread significantly along the circumference of the vortex core due to the weakness of the interaction.

The $\lambda_2$-criterion iso-surfaces \citep{jeong1995identification} plotted in figure \ref{fig:map_WI}(d)-(g) correspond to four different instants: slightly before (d), during (e), right after (f), and long after (g) the interaction. These instants are highlighted with vertical dashed lines in figures~\ref{fig:map_WI}(a)-(c). The vortex core structure can be visualised in these snapshots, showing that the ring remains highly coherent throughout the interaction, although it loses intensity. However, a remarkable loss of vorticity at the core centre can be observed by the reduction of opaque blue structures, although the vortex ring structure remains almost intact. Even though the interaction with the bubble is brief and localised, it triggers a strong decay of the vorticity in the core. These snapshots also confirm that the bubble (magenta surface) is attracted towards the vortex core but appears intact and unbroken upon the closest approach. This absence of rupture is consistent with the interaction being subcritical, as bubble breakup generally requires $We \gtrsim 1$ for $R_0/R_b > 1$~\citep{foronda2021deformation}. In fact, notice that the bubble covers a nearly constant azimuthal extension throughout the interaction (black solid lines in the maps of figures~\ref{fig:map_WI}a-c), indicating no remarkable shape deformations of the bubble surface. Note that a change of the extension of the azimuthal planes covered by the bubble could indicate not only a real shape alteration but also a radial motion of the bubble: for a given shape, the bubble covers a larger extension in the azimuthal coordinate when it moves to a smaller radial position, and \textit{vice versa} (see figure~\ref{fig:problem_sketch}b). In fact, the decrease that takes place prior to the interaction corresponds to the bubble motion towards the core (higher radial position), while the increase after the interaction corresponds to the migration away from the core towards the central vertical axis, in agreement with figure~\ref{fig:trajectory_WI}(c).\\

\begin{figure}
    \centering
    \includegraphics[width=\linewidth]{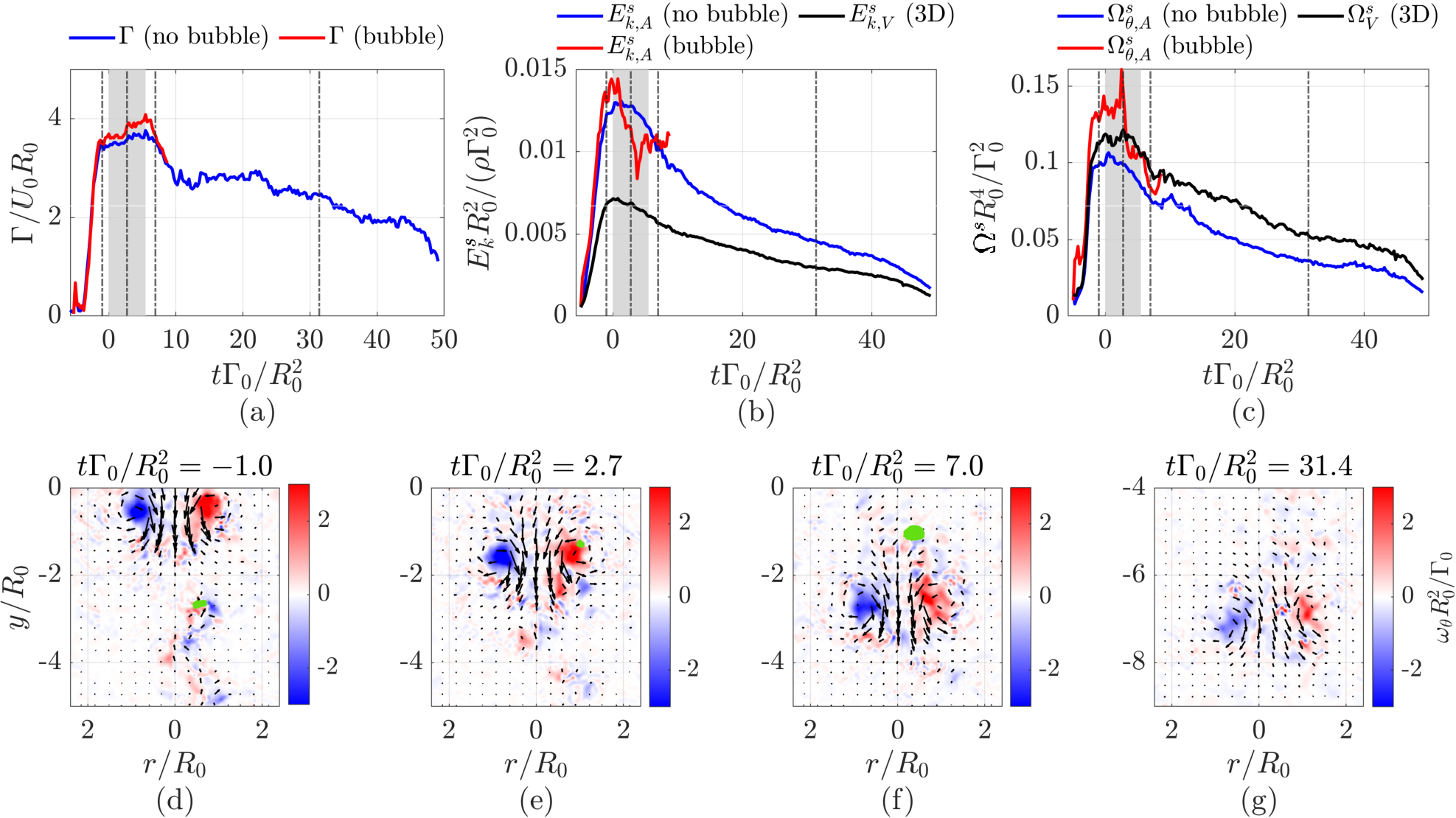}
    \caption{(Top row) Temporal evolution of dimensionless (a) circulation, (b) specific, kinetic energy, and (c) specific enstrophy in Case I (weak interaction). Results in planes with or without bubble are depicted in red and blue, respectively. Figures (b) and (c) show the azimuthal average of the corresponding 2D variable averaged in half-planes (red and blue) together with their 3D volumetric counterpart (black). The shaded areas correspond to the duration of the interaction. The black dotted lines mark the instants corresponding to different events throughout the experiment. (Bottom row) Velocity field and vorticity contours of the four instants marked in the top-row figures, on a vertical plane at $\theta=56^\circ$. The silhouette of the bubble is depicted in green. The bubble crosses this plane at all the depicted instants, but to a different extent depending on its azimuthal position and deformation at such instant (see the different sizes and shapes of the silhouettes in d-f).}
    \label{fig:dynamics_WI}
\end{figure}

The previous maps indicate that the bubble effect in this case is localised around the collision area. To further contrast the effect on the flow variables in this area ($\theta_{bubble}$) to the rest of the planes, where no bubble is present, we averaged them azimuthally, making a distinction between bubble and no-bubble azimuthal planes (red and blue lines, respectively in figures~\ref{fig:dynamics_WI}(a)-(c), calculated with equations~\ref{eq:eqs_averaged2D})\footnote{Note that, in these figures we only include the components of the velocity in the measuring plane to compute the kinetic energy and vorticity component perpendicular to the plane in the enstrophy, like in two-dimensional PIV measurements. However, since we perform three-dimensional measurements, the three components of the velocity and vorticity could have been obtained.}, and separating the information in half-planes ($\theta\in[0^\circ,360^\circ]$). Although the interaction is brief, notice that the bubble alters the vortex characteristics while it occurs (red line). Circulation is observed to be slightly enhanced during the interaction in the planes where the bubble is present, while significantly larger values of enstrophy and, mainly, kinetic energy are observed at the beginning of the interaction (red lines). As soon as the bubble starts moving away from the vortex core due to buoyancy, there is a quick drop of the enstrophy peak, reaching the value of the no-bubble planes (blue lines). At the same time, a decrease in energy happens, which may be understood as the bubble slowing down the vortex due to a wake effect. Thus, we can appreciate in the three graphs an evident change of slope to a steeper decay of the flow variables when the bubble is present, indicating that, despite the encounter being brief, the vortex loses strength due to its collision with the bubble. This effect, where the vortex is weakened by the interaction, aligns with findings from other studies, such as the experimental work by \citet{jha2015interaction}. Here, they reported a large reduction in enstrophy and a significant drop in the vortex convection speed in low-Weber-number cases where the bubble was briefly trapped, contrasting with negligible effects seen at higher $We$. This highlights the importance of performing measurements precisely in the azimuthal planes where the interaction occurs. This is feasible in three-dimensional experiments, where the interaction plane can be identified and selected afterwards. In this regard, the volume-averaged 3D value in the kinetic energy and enstrophy, equations~\eqref{eq:eqs_averaged3D}, are included in figures~\ref{fig:dynamics_WI}(b) and (c), respectively (black line). Note that the volumetric metrics integrate over the 3D fluid volume, accounting for all three components of velocity and, thus, vorticity. While planar magnitudes clearly show the impact of interaction with the bubble, three-dimensional magnitudes depict smoother evolution, although they display the changes in the kinetic energy and the total enstrophy when the bubble faces the vortex ring. Nevertheless, the decay of both magnitudes over time reflects the overall evolution of the vortex ring correctly.

The velocity field and vorticity of the four instants selected in figure \ref{fig:dynamics_WI} (a)-(c), along with the vorticity contours and the bubble surface, are shown in figures \ref{fig:dynamics_WI} (d)-(g). Note that we selected a plane containing the bubble at all times in order to characterise its effect. However, due to the azimuthal motion of the bubble over time, its presence may only be partial at certain instants, depending on whether the plane crosses it far from the centroid. As expected, high velocities are observed around the vortex core, while the liquid is almost stagnant far from the vortex ring. It can be observed that, in the pre-interaction and capture stages (figures \ref{fig:dynamics_WI} d and e, respectively), the vortex structure remains almost unchanged in terms of its shape and amount of vorticity. However, because of the bubble interaction, the vorticity of the core is heavily distorted locally once the bubble escapes (figure~\ref{fig:dynamics_WI} f). Finally, the vortex continues to translate downwards, but, although still coherent, with a weakened core (figure~\ref{fig:dynamics_WI} g). 
\subsection{Case II: Capture, drag and release}\label{subsec:caseII}
\begin{figure}
    \centering
    \includegraphics[width=\linewidth]{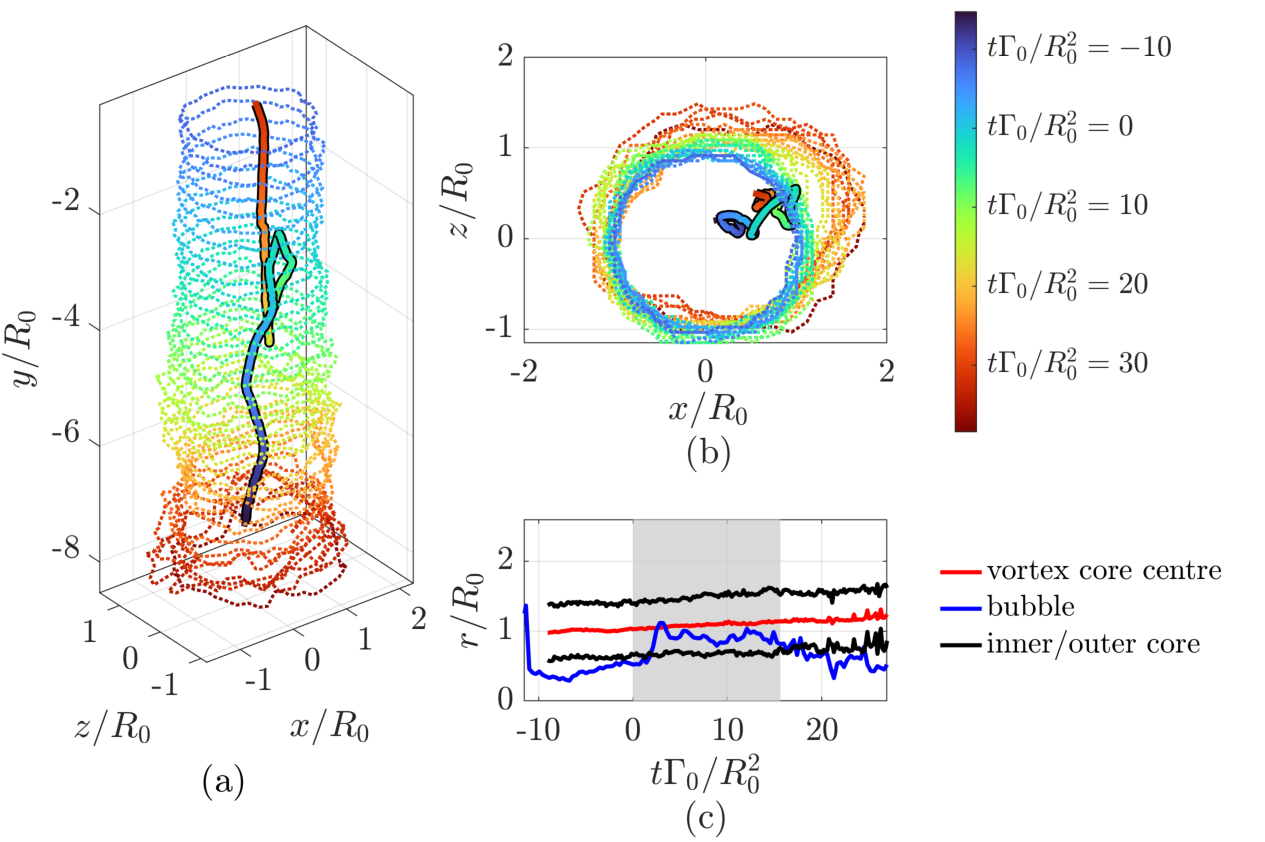}
    \caption{(a) Three-dimensional and (b) top views of the bubble trajectory (thick solid line) and time evolution of the shape and position of the vortex ring core centre (thin dotted lines) in the capture and drag case (Case II). Both are coloured by time (see colourbar), with the bubble travelling upwards and the vortex downwards (counterflow). (c) Temporal evolution of the azimuthally averaged radial position of the vortex core centre and inner/outer limits, together with the bubble centroid. The shaded area in panel (c) corresponds to the duration of the interaction.}
    \label{fig:trajectory_CDR}
\end{figure}
We will now describe a stronger interaction case, with higher $Re$ and $We$ than Case I, but similar vortex core-to-radius ratio, $a_0/R_0$. Figures~\ref{fig:trajectory_CDR}(a)-(b) describe the position of the vortex-bubble pair over time. As can be observed in the bubble path (thick time-coloured line in figures~\ref{fig:trajectory_CDR}a, b and blue line in c), the bubble travels upwards first and then is dragged laterally to larger radial positions when it is close enough to the vortex ($t=0$). During a certain time, the bubble is trapped inside the core (see figures~\ref{fig:trajectory_CDR}b, c) and travels downwards together with the vortex (see the period where the bubble path and the vortex contour are the same colour in figures~\ref{fig:trajectory_CDR}a and b). Meanwhile, the presence of the bubble weakens the vortex core, introducing shape distortions. When the vortex is no longer powerful enough to hold the bubble in its core, the bubble is released and rises again (figure~\ref{fig:trajectory_CDR}a). This interaction eventually leads to a destabilisation of the vortex ring structure. This is most evident in figures~\ref{fig:trajectory_CDR}(a) and (b), where at later times (reddish colours), the vortex no longer shows a circular shape. Additionally, in figure~\ref{fig:trajectory_CDR}(c), at the moment the bubble leaves the ring ($t_i\Gamma_0/R_0^2\approx15$), the position of the vortex core (shown by the red and black lines) exhibits some distortion due to the ring structure no longer being discernible. Note that the interaction time is significantly longer in this case than in Case I, due to the higher vortex circulation and lower pressure in the core.

\begin{figure}
    \centering
    \includegraphics[width=\linewidth]{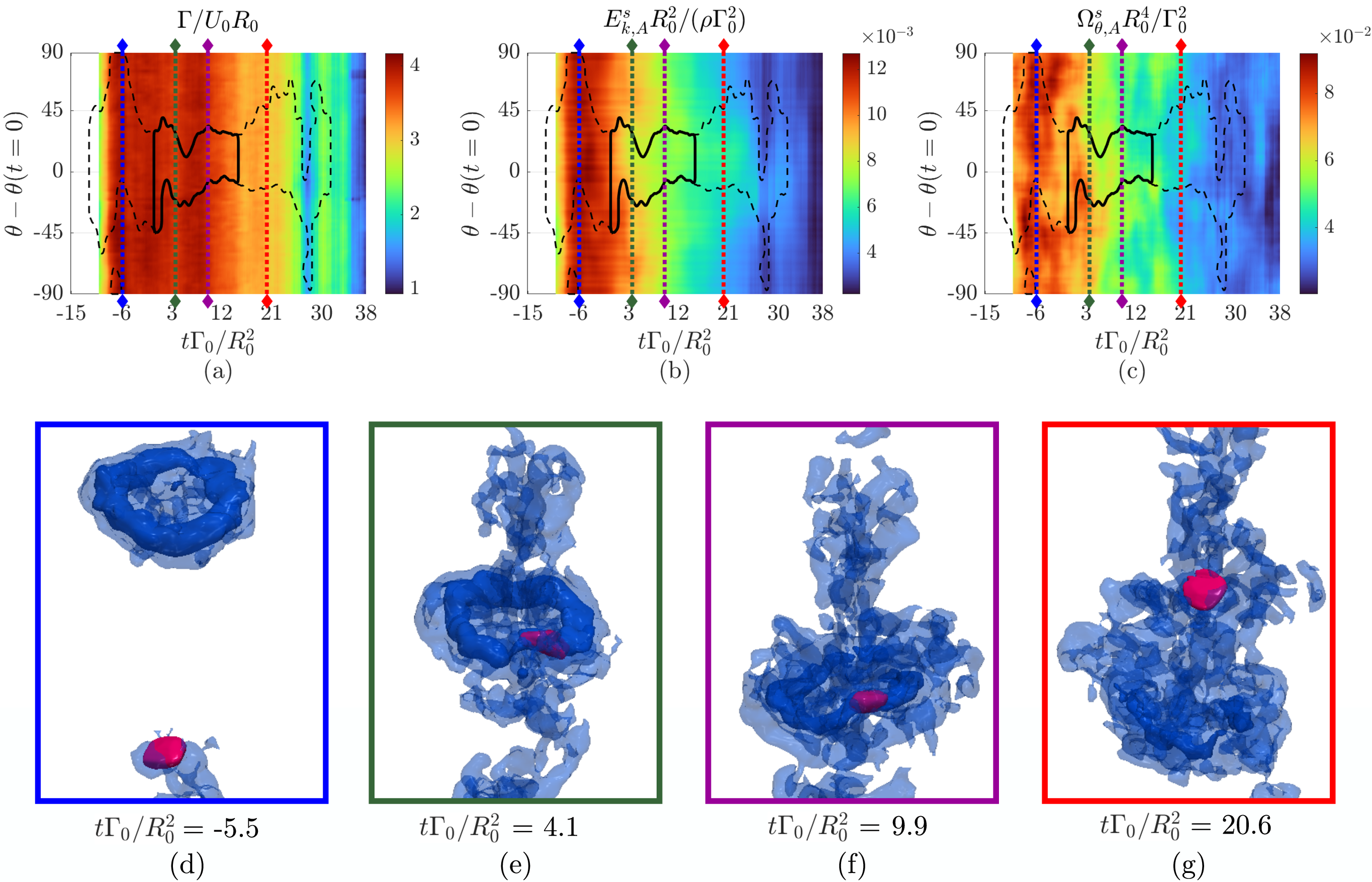}
    \caption{Temporal evolution of the azimuthal distribution of the (a) circulation, $\Gamma/(U_{0}R_{0})$, (b) planar, specific kinetic energy, $E_{k,A}^{s}R^{2}_{0}/(\rho \Gamma^{2}_{0})$, and (c) planar, specific azimuthal enstrophy, $\Omega_{\theta,A}^{s}R^{4}_{0}/\Gamma^{2}_{0}$, of the capture+drag+release case (Case II). The values correspond to planar azimuthal sections covering two cores with $\theta\in[0^\circ,180^\circ]$ (full planes), as it would be obtained in 2D-PIV measurements. The areas enclosed by black dashed lines mark the instants and azimuthal coordinates corresponding to the position of the bubble, $\theta_{bubble}$, while the solid ones correspond to its position while it remains inside the vortex core during the interaction. The dotted vertical lines indicate the instants corresponding to different snapshots throughout the experiment (d to g, matched by colour) for a better understanding of the effect of the interaction. The bubble surface is coloured in magenta, while blue represents the $\lambda_2$-criterion iso-surfaces (opaque blue: $\lambda_2 R^{4}_{0}/\Gamma^{2}_{0}=-0.72$; faded blue: $\lambda_2 R^{4}_{0}/\Gamma^{2}_{0}=-0.0036$). A supplementary movie related to this figure (Movie6\_caseII.mp4) is accessible at journals.cambridge.org.}
    \label{fig:map_CDR}
\end{figure}

The evolution of circulation, planar, specific energy and planar, specific enstrophy are depicted in figure~\ref{fig:map_CDR}, together with the bubble position (black line enclosed area). As in Case I, circulation decreases over time, with no notable effects of the bubble during the initial stage of the interaction. However, after some contact between the ring and the bubble ($t\Gamma_0/R_0^2\gtrsim10$), the circulation starts to decrease more rapidly. This effect is even more evident in kinetic energy and enstrophy (figures~\ref{fig:map_CDR}b and c, respectively), with a steep decrease in their magnitude, which started much earlier ($t\Gamma_0/R_0^2\approx3$). Unlike in the previous case, it is also notable that the effects on the vortex ring are not limited to the vicinity of the bubble, but extend more globally around $\theta$. 
The value of the Weber number ($We=1$) is not high enough to cause the bubble to break up. This may be due to the thickness of the vortex ($a_0/R_0=0.38$), which makes it weaker than a thinner vortex with the same $Re$. Note that the result of the interaction and some differences between the cases may also be caused by factors not included in the scope of this work, such as the relative position of the bubble and the vortex at the beginning of the interaction.

Regarding the bubble position, before the interaction (dashed lines in figures~\ref{fig:map_CDR}a-c), we can see the bubble drifting back and forth from its initial position (covering different azimuthal extension), indicating it has an unstable path (see also figure~\ref{fig:trajectory_CDR}a). Right after that (see the change from the dashed to the solid line), the bubble suffers an attraction towards the vortex core at a higher radial distance (see also figure~\ref{fig:trajectory_CDR}c), covering thus a lower azimuthal extension. Once inside the core, the bubble drifts azimuthally and then travels back closer to $\Delta\theta\approx0$, where it stays until its release. This may be a consequence of the vortex ring losing intensity and coherence, as can be seen in figures~\ref{fig:map_CDR}(d)-(f) through the $\lambda_2$ criterion at four different instants. During capture and the initial bubble drag phase  (figures~\ref{fig:map_CDR}d-e), the vortex structure retains its overall shape and level of vorticity. This period marks the beginning of intense interactions that strongly affect instantaneous liquid vortex properties. During the sustained drag (figure~\ref{fig:map_CDR}f), the shape and symmetry of the ring are distorted due to the presence of the trapped bubble, in agreement with figures~\ref{fig:trajectory_CDR}(a)-(b). This distortion, while not yet causing major global circulation loss, reflects the strong perturbation introduced by the bubble. Figure \ref{fig:map_CDR}(g) corresponds to the post-release stage, during which the vortex core loses overall strength and coherence due to structural weakening. Regarding the bubble shape (magenta), note that although it is initially deformed at the beginning of the interaction due to the stresses induced by the vortex, they are not sufficiently high to induce its breakup.
\begin{figure}
    \centering
    \includegraphics[width=\linewidth]{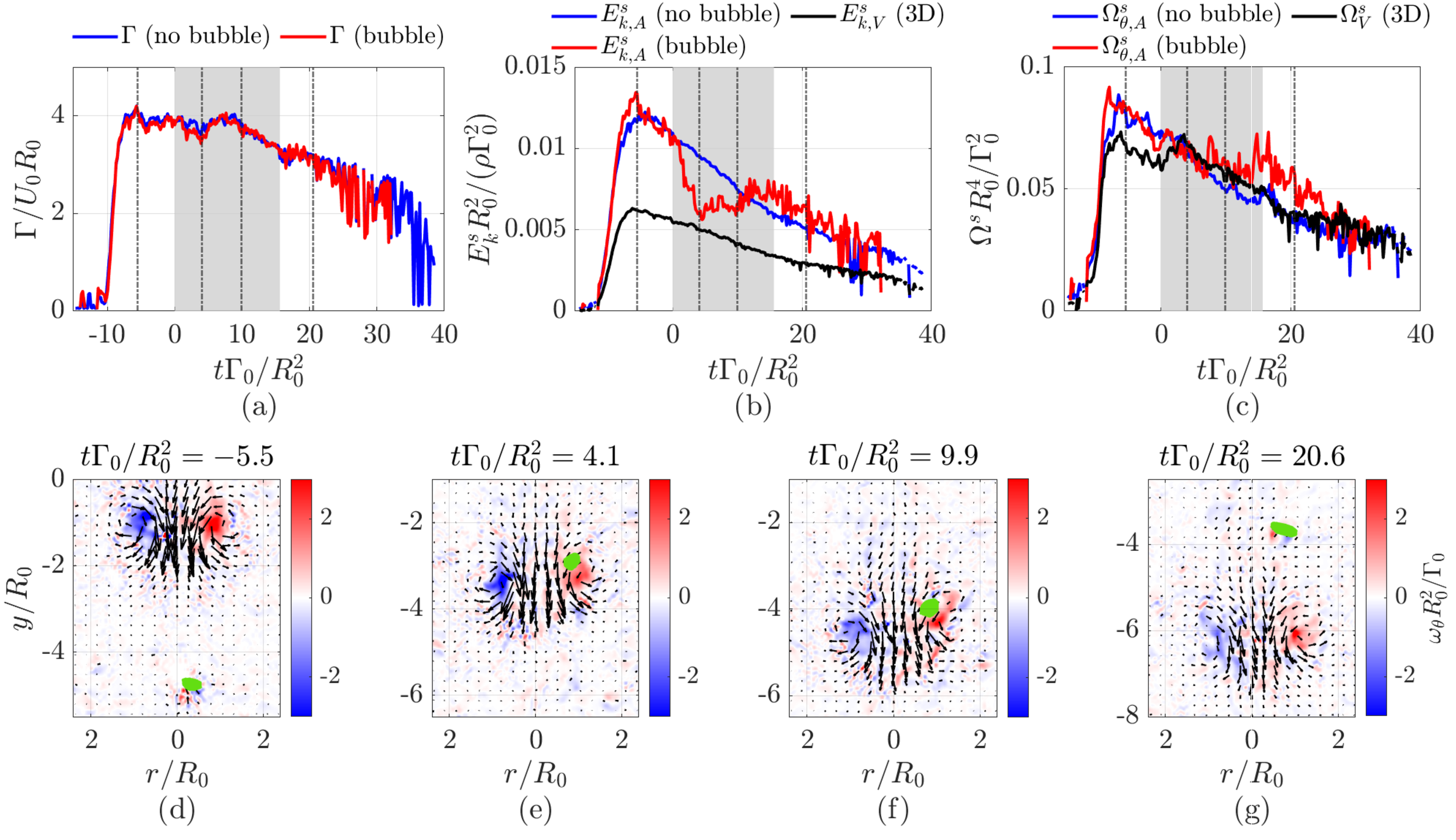}
    \caption{(Top row) Temporal evolution of dimensionless (a) circulation, (b) specific, kinetic energy, and (c) specific, azimuthal enstrophy in Case II (capture+drag+release). Results in planes with or without bubble are depicted in red and blue, respectively. Figures (b) and (c) show the azimuthal average of the corresponding 2D variable averaged in half-planes (red and blue) together with their 3D volumetric counterpart (black). The shaded areas correspond to the duration of the interaction. The black dotted lines mark the instants corresponding to different events throughout the experiment. (Bottom row) Velocity field and vorticity contours of the four instants marked in the top-row figures, on a vertical plane at $\theta=61^\circ$. The silhouette of the bubble is depicted in green. The bubble crosses this plane at all the depicted instants, but to a different extent depending on its azimuthal position and deformation at such instant (see the different sizes and shapes of the silhouettes in d-g).}
    \label{fig:dynamics_CDR}
\end{figure}

To further analyse how the position of the bubble impacts the flow over time, we may look at figure~\ref{fig:dynamics_CDR}. Here, we cannot identify any significant differences in the decay of circulation over time between planes with and without bubble (figure~\ref{fig:dynamics_CDR}a), which is consistent with the findings by \citet{jha2015interaction} for high enough Weber numbers and thick core vortices. However, a slight increase in circulation can be observed around $t\Gamma_0/R_0^2 \approx 6$, probably due to the vorticity shed by the bubble when it starts to interact with the vortex ring. In contrast, we see a strong effect of the bubble in kinetic energy and azimuthal enstrophy. In the former (figure~\ref{fig:dynamics_CDR}b), both the 3D evolution and the mean value of the 2D half-planes without bubble have a similar trend: a monotonic decrease over time, with a slightly steeper slope during the interaction period. However, the planes where the bubble interacts with the vortex show a significant loss of kinetic energy from the start of the interaction. This is a direct consequence of the perturbations introduced by the presence of the bubble, slowing down the vortex. In fact, figures~\ref{fig:dynamics_CDR}(e) and (f) show how the vortex tilts with the core on the left side lower than that on the right, with the latter containing the bubble. Shortly before the bubble escapes from the vortex, the two-dimensional specific kinetic energy in the planes containing the bubble recovers to a value similar to that of the planes without bubble, exhibiting a bump (figure~\ref{fig:dynamics_CDR}b). Note that the rate of decay of the three-dimensional, specific kinetic energy is also higher during interaction, which is consistent with direct numerical simulations by \citet{foronda2021deformation}. Regarding the time evolution of the azimuthal enstrophy, shown in figure~\ref{fig:dynamics_CDR}(c), different trends are observed. In the planes containing the bubble, it initially decreases when the bubble starts to interact with the vortex ring. However, during this interaction, $\Omega^s_{\theta,A}$ remains almost constant for a period of time before increasing as the bubble escapes from the vortex. This increase coincides with the rise of $E^s_{k,A}$ shown in figure~\ref{fig:dynamics_CDR}(b). Differently, in planes which do not contain the bubble, a monotonic decrease of $\Omega^s_{\theta,A}$ is observed with time (blue line in figure~\ref{fig:dynamics_CDR}c), with a slightly steeper slope during the interaction period. However, the volumetric enstrophy (black line in figure~\ref{fig:dynamics_CDR}c) shows an increase at the beginning of the interaction, in agreement with the 3D numerical simulations by \citet{foronda2021deformation}. 

Finally, the evolution of the vortex and how it loses energy can be further observed in the velocity and vorticity fields in a plane containing the bubble, represented in figures~\ref{fig:dynamics_CDR}(d)-(g). Similarly to the previous case, here the bubble quickly approaches the ring and gets trapped (figures~\ref{fig:dynamics_CDR}d-e). Unlike in the weak interaction case, instead of rapidly escaping, it is dragged downwards. In this interplay, the vortex level of vorticity is almost unaltered (figure~\ref{fig:dynamics_CDR}f), but the shape and symmetry of the ring are heavily distorted. It is not until the bubble escapes (figure~\ref{fig:dynamics_CDR}g) that the level of vorticity is importantly diminished. 
In summary, this case depicts a combination of local effects in the bubble surroundings, which then evolve to a strong global distortion of the vortex due to the presence of the bubble.

\subsection{Case III: Bubble breakup}\label{subsec:caseIII}
\begin{figure}
    \centering
    \includegraphics[width=\linewidth]{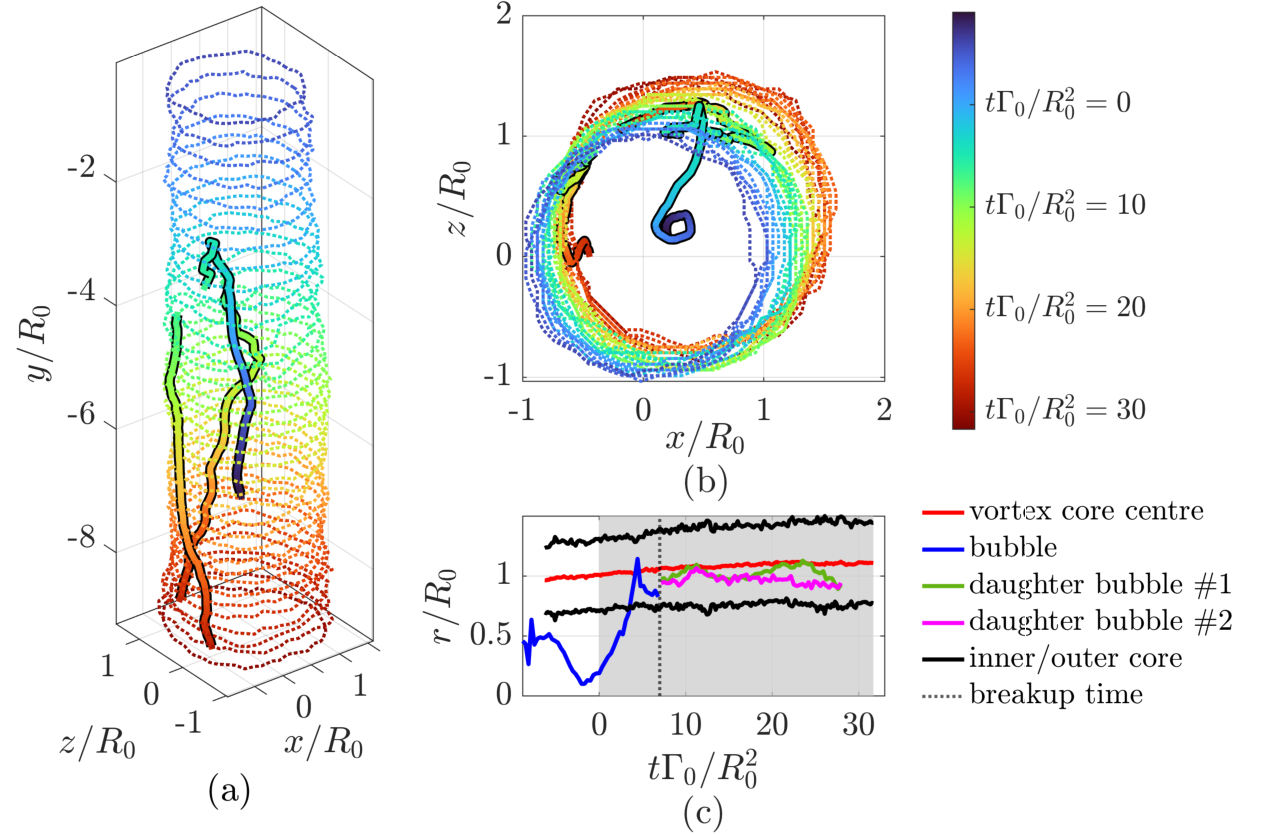}
    \caption{(a) Three-dimensional and (b) top views of the bubble trajectory (thick solid line) and time evolution of the shape and position of the vortex ring core centre (thin dotted lines) in the breakup case (Case III). Both are coloured by time (see colourbar), with the bubble travelling upwards and the vortex downwards (counterflow). (c) Temporal evolution of the azimuthally averaged radial position of the vortex core centre and inner/outer limits, together with the bubble centroid. The shaded area in panel (c) corresponds to the duration of the interaction.}
    \label{fig:trajectory_BBB}
\end{figure}
The last case in this work covers the strongest interaction: here, the vortex ring breaks the bubble into two daughter bubbles. The kinematics of this interaction are depicted in figure~\ref{fig:trajectory_BBB}. The three-dimensional and top views show that initially, the bubble rises near the vertical axis of the vortex ring while the ring travels downwards (dark to light blue in figures~\ref{fig:trajectory_BBB} a, b). When they are close enough, the bubble is quickly deviated towards the ring core at $t=0$ (see also figure~\ref{fig:trajectory_BBB}c), and starts to travel downwards within the vortex core. In this case, the vortex is thinner than in Cases I and II, and although $We$ is smaller than in Case II, it is large enough to overcome the confining surface tension forces acting on the bubble. Thus, after a very short time, the mother bubble splits into two daughter bubbles, which move in the azimuthal direction without leaving the vortex core for the duration of the experiment. Regarding the vortex ring, its circular shape remains largely unaltered, as shown in figures~\ref{fig:trajectory_BBB}(a)-(b).
\begin{figure}
    \centering
    \includegraphics[width=\linewidth]{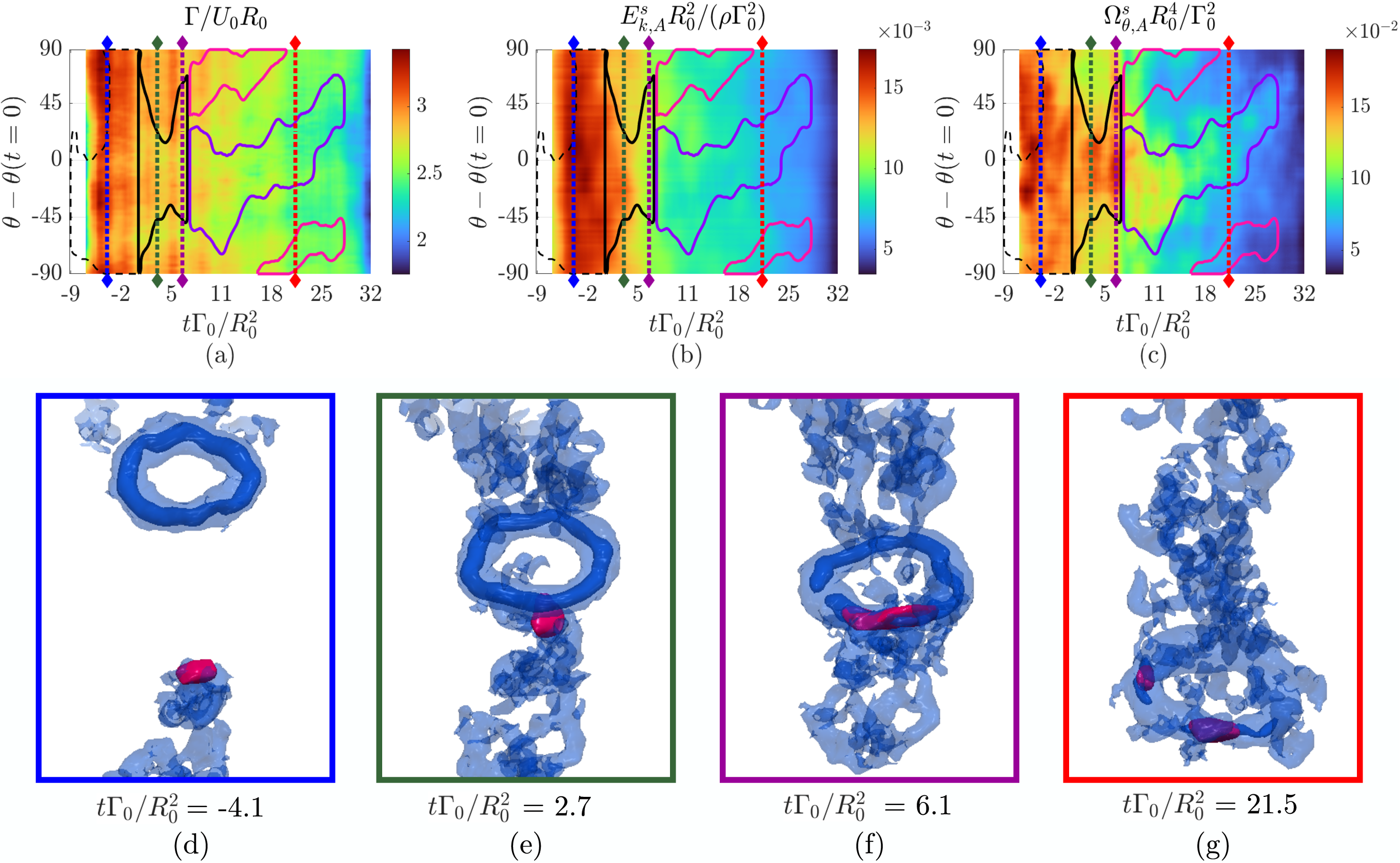}
    \caption{Temporal evolution of the azimuthal distribution of the (a) circulation, $\Gamma/(U_{0}R_{0})$, (b) planar, specific kinetic energy, $E_{k,A}^{s}R^{2}_{0}/(\rho \Gamma^{2}_{0})$, and (c) planar, specific azimuthal enstrophy, $\Omega_{\theta,A}^{s}R^{4}_{0}/\Gamma^{2}_{0}$, of the breakup case (Case III). The values correspond to planar azimuthal sections covering two cores with $\theta\in[0^\circ,180^\circ]$ (full planes), as it would be obtained in 2D-PIV measurements. The areas enclosed by black dashed lines mark the instants and azimuthal coordinates corresponding to the position of the mother bubble, while the solid ones correspond to its position while it remains inside the vortex core during the interaction. After the mother bubble breaks, each of the daughter bubbles is marked with a different colour of the solid line (large bubble in purple and small bubble in magenta). The dotted vertical lines indicate the instants corresponding to different snapshots throughout the experiment (d to g, matched by colour) for a better understanding of the effect of the interaction. The bubble surface is coloured in magenta, while blue represents the $\lambda_2$-criterion iso-surfaces (opaque blue: $\lambda_2 R^{4}_{0}/\Gamma^{2}_{0}=-3.80$; faded blue: $\lambda_2 R^{4}_{0}/\Gamma^{2}_{0}=-0.0038$). A supplementary movie related to this figure (Movie7\_caseIII.mp4) is accessible at journals.cambridge.org.}
    \label{fig:map_BBB}
\end{figure}

Figure~\ref{fig:map_BBB} shows how the flow variables evolve with the azimuthal position of the bubble. Similarly to the previous cases, the evolution of the flow variables shown in figures~\ref{fig:map_BBB}(a)-(c) is characterised by a monotonic decay over time, which accelerates when the interaction begins. No particular effects on the temporal evolution of the circulation are spotted in figure~\ref{fig:map_BBB}(a), only slight changes appear along the azimuthal direction due to the perturbations induced by the bubbles. The kinetic energy (figure~\ref{fig:map_BBB}b) shows higher values in the area around the bubble at the moment of capture. After this peak, the value quickly stabilises globally around the azimuth. Finally, the azimuthal enstrophy (figure~\ref{fig:map_BBB}c) is much more strongly affected by the presence of the bubble, particularly during its breakup. Before the interaction, it displays a nearly uniform distribution around $\theta$. However, when the interaction begins, it concentrates at the area of the bubble, while decreasing in the surroundings. It even increases and peaks a few instants before breakup, after which it quickly drops. This increase in enstrophy is associated with the bubble elongation within the vortex core. Moreover, the enstrophy remains higher around the daughter bubbles than elsewhere in the ring throughout the experiment, particularly in the area of the largest daughter bubble (purple solid line in figure~\ref{fig:map_BBB}c). This indicates that even though the vortex structure in this case remains coherent, as can be noticed in figures~\ref{fig:map_BBB}(d)-(g), and thus less affected by the bubble, it gets perturbed, shedding vorticity, especially due to large bubbles trapped in the core (see figure~\ref{fig:map_BBB}g). 
\begin{figure}
    \centering
    \includegraphics[width=\linewidth]{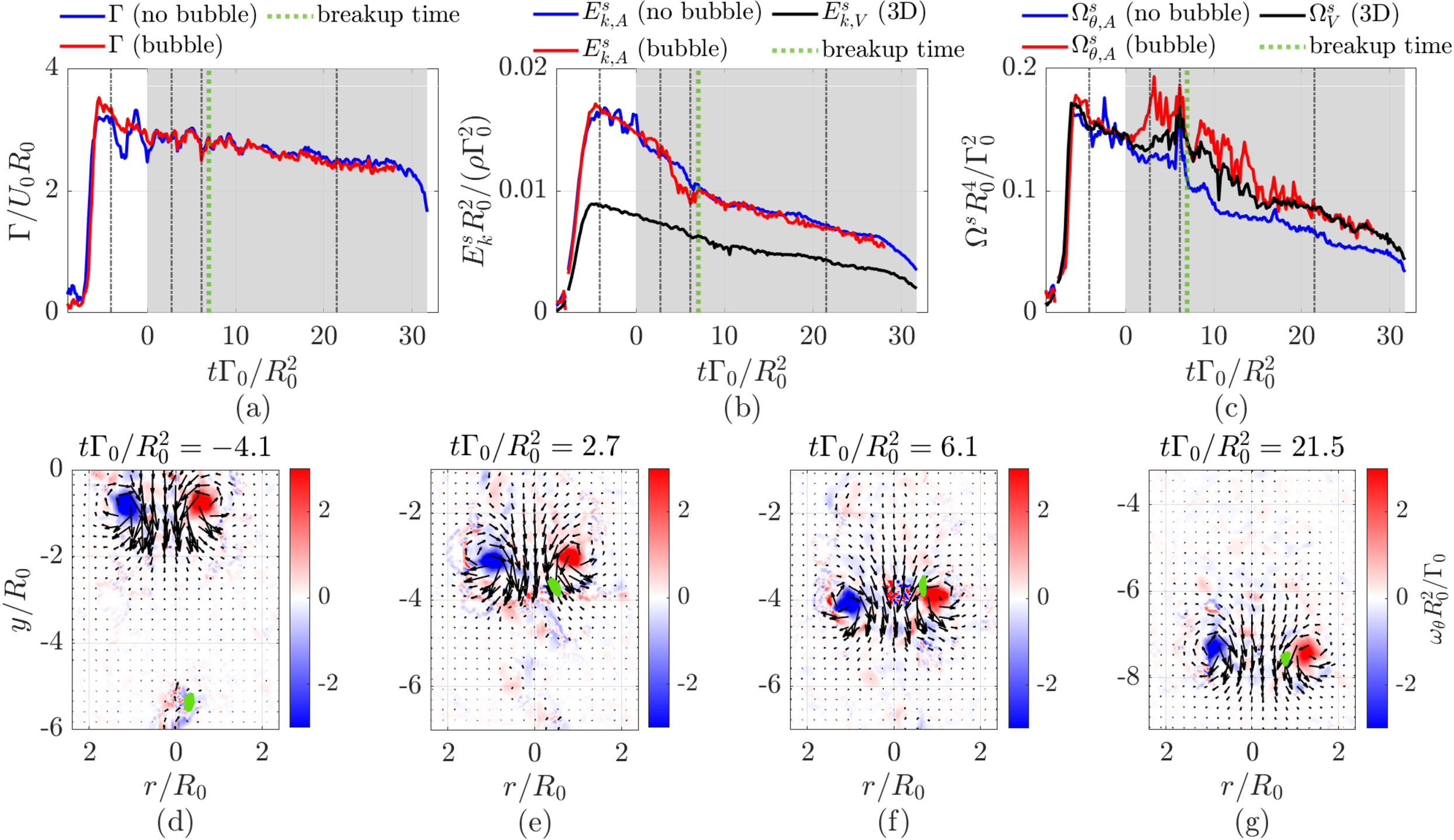}
    \caption{(Top row) Temporal evolution of dimensionless (a) circulation, (b) specific, kinetic energy, and (c)  specific, azimuthal enstrophy in the Case III (bubble breakup). Results in planes with or without bubble are depicted in red and blue, respectively. Figures (b) and (c) show the azimuthal average of the corresponding 2D variable averaged in half-planes (red and blue) together with their 3D volumetric counterpart (black). The shaded areas correspond to the duration of the interaction. The black dotted lines mark the instants corresponding to different events throughout the experiment, and the green dotted line marks the breakup time. (Bottom row) Velocity field and vorticity contours of the four instants marked in the top-row figures, on a vertical plane at $\theta=25^\circ$. The silhouette of the bubble is depicted in green. The bubble crosses this plane at all the depicted instants, but to a different extent depending on its azimuthal position and deformation at such instant (see the different sizes and shapes of the silhouettes in d-g).}
    \label{fig:dynamics_BBB}
\end{figure}
If we focus now on the bubble dynamics, it shows a motion around the central axis before the interaction (black dashed line in figures \ref{fig:map_BBB}a-c): initially, the bubble covers all the $\Delta\theta$ values because it is at the centre of the axial axis (see also figure~\ref{fig:map_BBB}d). A few instants later (figure~\ref{fig:map_BBB}e), the bubble rises to a vertical position a few bubble diameters below the vortex ring, where it starts to be influenced by its velocity. At $t=0$, it is quickly dragged laterally into the vortex core (see also figure~\ref{fig:trajectory_BBB}c). In fact, note that the azimuthal area covered by the bubble (solid black line in figures~\ref{fig:map_BBB}(a)-(c) rapidly decreases to a minimum of $\theta_{bubble}\sim90^\circ$. It is after this minimum in the spread of the bubble in $\theta$ (see figure \ref{fig:map_BBB}e) that it grows again due to an azimuthal elongation of the bubble along the ring core (see figure \ref{fig:map_BBB}f). The sequence of capture, elongation, and subsequent breakup observed here aligns closely with the description given by \citep{jha2015interaction}, who detailed a four-stage interaction process in which a bubble is attracted to the low-pressure core, elongates due to azimuthal pressure gradients, and eventually breaks. The elongation is a consequence of the bubble perturbing the pressure distribution inside the core: the pressure rises at the location of the bubble, creating a pressure difference along the azimuthal direction that causes the bubble to expand along $\theta$~\citep{martinez2015bubbles,biswas2022interaction,biswas2023vortex,zhang2023modification}, covering approximately up to one third of the circumference of the ring (see figure~\ref{fig:map_BBB}f). The elongation stage is very short, because the bubble quickly breaks into two parts (see purple and magenta solid lines in figures~\ref{fig:map_BBB}(a)-(c) at $t\Gamma_0/R_0^2\simeq6$). These daughter bubbles have different sizes, and their spread in $\theta$ tends to reduce after breakup, given that the surface tension is strong enough to restore the spherical-to-ellipsoidal shape of the bubbles (see figure~\ref{fig:map_BBB}g). They also exhibit a drift along $\theta$, like in the interaction of a vortex ring with a solid particle \citep{biswas2025interaction}.

Now, we summarise the effect of the bubble on the flow characteristics by examining how the averaged values of kinetic energy and enstrophy differ in half-planes where the bubble interacts with vortex motion, compared to half-planes free of interaction, alongside their three-dimensional counterparts. Figure~\ref{fig:dynamics_BBB}(a) shows that, as seen before, circulation is hardly affected by the bubble, with a moderate decay in time. The planar, specific kinetic energy (see figure~\ref{fig:dynamics_BBB}b) also decays, in both planes with and without the bubble. However, it undergoes a significant change in slope a few moments after the bubble breaks up (marked by a green dotted line in figures~\ref{fig:dynamics_BBB}a-c). In the planes with a bubble (red line), the energy is redistributed near the bubble, which results in a slight increase in energy, which quickly drops before breakup and then recovers to the same value as the rest of the ring, in agreement with figure~\ref{fig:map_BBB}(b). Note also the rapid decrease of $E^s_{k,A}$ around $t\Gamma_0/R^2_0 \approx 2.7$ in the planes containing the bubble. By looking at the snapshots in figures~\ref{fig:dynamics_BBB}(d)-(f), we may observe that the liquid velocity (and thus the kinetic energy) does not show a strong difference when the bubble starts to enter the vortex (figures~\ref{fig:dynamics_BBB}d-e). However, the values of both velocity and energy are much smaller when the bubble is in its equilibrium radial position inside the core for a sufficiently long time. (figure~\ref{fig:dynamics_BBB}f). Similarly to the previous cases, the volumetric kinetic energy is reduced, and its temporal decay is similar but smoother than that of its planar counterpart, although it also shows the change of slope after the bubble breakup. 

Finally, regarding enstrophy (figure~\ref{fig:dynamics_BBB}c), there is a significant increase from the moment the bubble is attracted by the vortex at $t=0$ (see also figure~\ref{fig:dynamics_BBB}e). This increase in enstrophy is associated with the kinetic energy decrease observed at around $t\Gamma_0/R^2_0 \approx 2.7$. The growth in enstrophy is fairly maintained until the bubble reaches its maximum elongation (see figure~\ref{fig:dynamics_BBB}f). Afterwards, the enstrophy decreases. Following the breakup, the enstrophy in the bubble half-planes (red line) begins to decay more quickly than before the bubble breakup. This decay is even more pronounced in the no-bubble planes (blue line). This indicates not only the growth of vorticity near the bubbles, but also its redistribution from the rest of the ring towards the bubble areas, as shown on the map in figure~\ref{fig:map_BBB}(c). In figure~\ref{fig:dynamics_BBB}(g), the level of vorticity has diminished notably after breakup. Regarding the evolution of three-dimensional enstrophy, its behaviour is a combination of planar enstrophy in planes with and without a bubble. It also shows an increase in enstrophy when the vortex ring traps the bubble. It is worth mentioning that the evolution of kinetic energy and azimuthal enstrophy in Cases II and III are very much alike, which may be expected due to their similar $Re$ and $R_0/R_b$ values and relatively high $We$. The main difference between them is $a_0/R_b$, with Case III presenting a thinner core, which indicates that core thickness is a key feature to enable bubble breakup. \\

The results reported in this Section show that two-dimensional measurements are often insufficient to capture the complex, transient, and nonlinear three-dimensional dynamics governing these phenomena. Although classical two-dimensional particle image velocimetry (2D-PIV) can be used to investigate the interaction between a rising bubble and a vortex ring, its ability to fully characterise the flow is inherently limited, particularly when the interaction induces out-of-plane motion and a loss of axial symmetry. Moreover, the use of a fixed measurement plane restricts the observation of spatially evolving features of the interaction. The present results show that even weak interactions can produce measurable changes in the global structure of the vortex, which may not be fully captured by planar approaches, highlighting an important limitation of such techniques. 
Thus, the fixed measurement plane in 2D-PIV must coincide with the interaction region, a condition that is difficult to ensure in practice due to the continuous displacement of the bubble during its rise. As a result, the capture of such localised phenomena becomes highly sensitive to the positioning of the measurement plane. In contrast, time-resolved three-dimensional Lagrangian particle tracking (4D-LPT), coupled with shadowgraphy, enables a volumetric characterisation of the flow within the measurement domain. This approach allows for the simultaneous tracking of three-dimensional bubble deformation and the reconstruction of the three components of the velocity field, from which the vorticity field can be estimated. Furthermore, it enables \textit{a posteriori} identification of the interaction region and provides a more comprehensive description of the temporal evolution of global quantities such as total kinetic energy and enstrophy. Consequently, this methodology alleviates key limitations of planar measurements and enables a more comprehensive assessment of whether vorticity is dissipated or redistributed following the loss of axial symmetry, as well as capturing the inherently three-dimensional nature of vortex instabilities and fragmentation processes.

\FloatBarrier
\subsection{Comparative analysis}
\begin{figure}
    \centering
    \includegraphics[width=\linewidth]{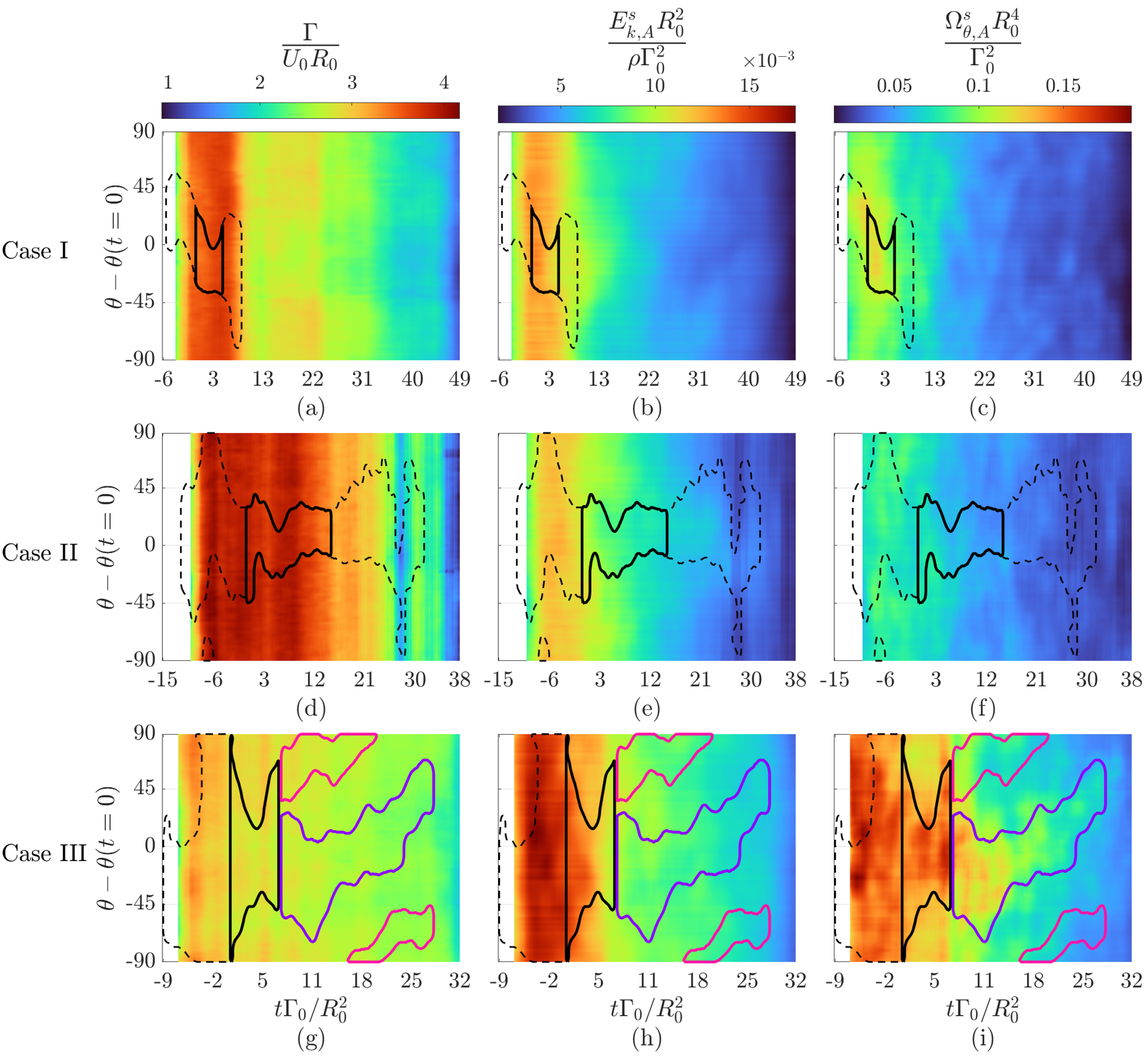}
    \caption{Temporal evolution of the azimuthal distribution of the dimensionless circulation (left column), planar specific kinetic energy (central column), and planar specific azimuthal enstrophy (right column) for Case I (first row), Case II (second row), and Case III (third row).}
    \label{fig:compare_maps}
\end{figure}
Figure \ref{fig:compare_maps} directly compares the latter magnitudes in the $\Delta \theta-t$ planes for three interaction cases. If we focus on circulation (left column), Case II figure~\ref{fig:compare_maps}(d) shows larger values compared to Cases I and III (figures~\ref{fig:compare_maps} a and g, respectively), which can be attributed to the initial strength of the vortex generated for that experiment. In fact, Case II was intentionally set up with the highest circulation strength, resulting in the largest Reynolds number among the three scenarios. Although all magnitudes decay over time, circulation is the magnitude least affected by the bubble interaction. This means that the ranking established by the initial vortex strength (Case II $>$ Case III $>$ Case I) is maintained throughout the measurement time. 

Nevertheless, the planar, specific kinetic energy level (central column of  figure~\ref{fig:compare_maps}) is much higher in Case III (figure~\ref{fig:compare_maps}h) compared to Cases I and II (figure~\ref{fig:compare_maps} b and e, respectively) primarily because Case III involves an interaction with a vortex ring that has initially a higher kinetic energy (see Table \ref{tab:initial}) and a thinner core, which concentrates the flow variables more intensely. Moreover, kinetic energy in Case III (figure~\ref{fig:compare_maps}h) shows higher values in the area around the bubble at the moment of capture. This greater kinetic energy is related to the vorticity emitted by the bubble during the capture process, which consists first in stopping the bubble rise and then being directed towards the bubble core (see figure~\ref{fig:map_BBB}e). In this case, the bubble changes its velocity more abruptly, which could explain the raise of the kinetic energy. After this initial peak, the kinetic energy magnitude quickly stabilises globally around the azimuth, maintaining a high level compared to the rapidly decaying energy found in the strongly distorted vortex core of Case II (figure~\ref{fig:compare_maps}e).

Levels of specific azimuthal enstrophy (right column of  figure~\ref{fig:compare_maps}) are observed to be the highest also in Case III (figure~\ref{fig:compare_maps}i), compared to Cases I and II (figure~\ref{fig:compare_maps} c and f, respectively). This fact is mainly due to the specific vortex properties of Case III and the mechanism of intense vorticity concentration and stretching leading up to the bubble rupture. A thinner core structure means that the vortex principal vorticity is concentrated in a smaller cross-sectional area~\citep{biswas2023vortex}. This concentration leads to higher vorticity (and enstrophy) when measured in a cross-sectional plane. In contrast, only a brief, localised perturbation occurs in Case I (figure~\ref{fig:compare_maps}c). In Case II (figure~\ref{fig:compare_maps}f), the azimuthal enstrophy is lower because the effect of the thin core is absent, what spreads the flow variables over a larger area. The expected influence of the vortex core can be accounted for by redefining the Weber number to include the effect of its size on the characteristic velocity scale. Thus, since for a vortex ring the velocity far from the bubble is given by $U_0= \Gamma_0/(4\pi R_0)ln[8R_0/a_0- 1/4]$~\citep{lamb1924hydrodynamics}, a characteristic velocity of the form $(\Gamma_0/4 \pi R_0) \, ln(8 R_0/a_0)$ can be used to define the Weber number, as in \citet{higuera2004axisymmetric}, $We^* = We/4 \, [ln(8 R_0/a_0)]^2$. According to the above definition of the Weber number, the values for Cases I, II, and III are 0.97, 2.32, and 2.16, respectively. This indicates that the effective Weber numbers in Cases II and III are very similar, and that the final outcome (trapping or breakup) is associated with the distance between the bubble and the vortex core at the onset of their interaction.

\begin{figure}
    \centering
    \includegraphics[width=\linewidth]{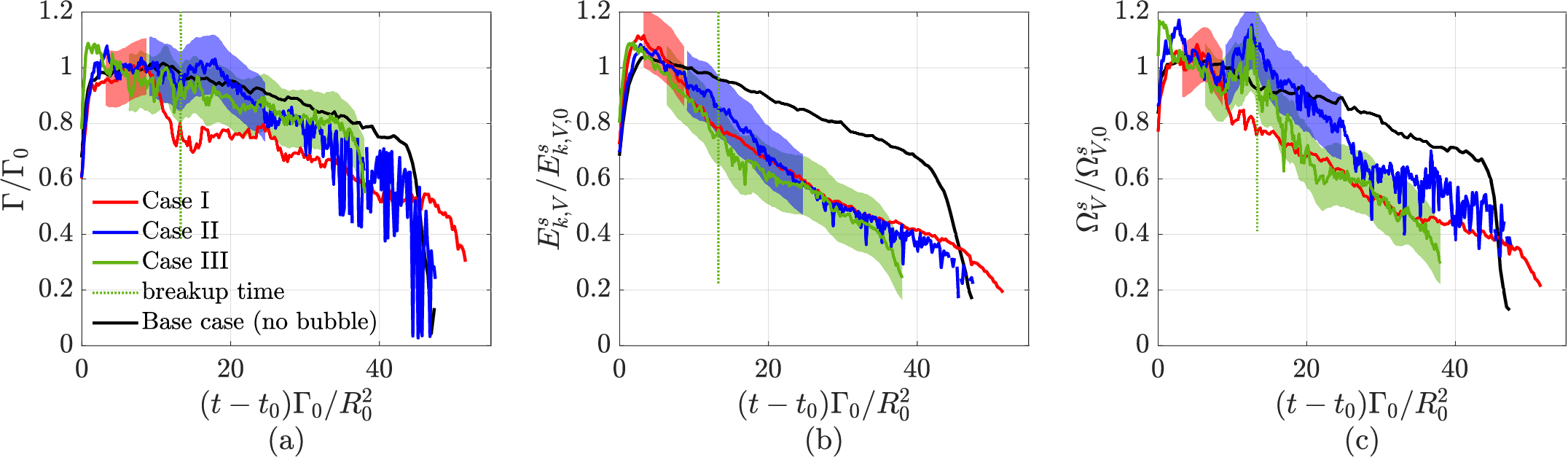}
    \caption{Evolution of (a) circulation, (b) specific, volumetric, kinetic energy, and (c) specific, volumetric, total enstrophy over time for a base case of vortex ring without bubble ($Re=28000$, $a_0/R_0=0.29$) and the different interaction cases studied: Case I (weak interaction), Case II (capture, drag, release) and Case III (bubble breakup). Coloured areas indicate interaction intervals. The vertical green dotted line denotes breakup time in Case III. Note that to compare the evolution of the different regimes, the flow variables are normalised to their initial value and the time scales start at $t-t_0$, with $t_0$ the initial time of the experiment, when the vortex has completely entered our measurement volume.}
    \label{fig:compare_dynamics}
\end{figure}

The three interaction cases show distinct responses that may be defined in terms of variations in the vortex volumetric ring magnitudes. In this regard, figure~\ref{fig:compare_dynamics} compares the evolution of (a) circulation ($\Gamma$), (b) specific volumetric kinetic energy ($E_{k,V}^s$), and (c) specific total volumetric enstrophy ($\Omega_V^s)$ for the base vortex ring (without a bubble) and the three distinct interaction cases. The duration of the kinematic interaction is also highlighted in figure~\ref{fig:compare_dynamics} through the shaded areas. As explained before, the volumetric magnitudes reflect the overall average evolution and long-term coherence of the vortex ring. Note that circulation (figure~\ref{fig:compare_dynamics}a) is the integral property least affected by the presence and interaction of the bubble, with the decay over time similar for all cases. However, in Case I, in which the vortex does not have enough energy to attract and disturb the bubble, the circulation decreases abruptly after the interaction due to the disturbance introduced by the bubble.

Regarding kinetic energy (figure~\ref{fig:compare_dynamics}b), all interaction cases exhibit an enhanced decay rate when compared to the base case, signifying that the bubble interaction leads to a general loss of strength in the vortex. In Case I, the interaction is characterised by a brief encounter with the bubble. However, the vortex still loses strength, resulting in a steeper decay slope than in the base case. This aligns with the observation that cases with a low Weber number result in overall weakening.~\citep{jha2015interaction}. The decay rate for Cases II and III is larger during the interaction, and it slows down after release or breakup, respectively, suggesting a significant loss of kinetic energy from the beginning of the capture process. 

Volumetric total enstrophy ($\Omega_V^s$) is the integral property most strongly affected by the bubble interaction. Notice that, in the three cases, $\Omega^s_V$ increases when the vortex starts to interact with the bubble, showing a bump in the time evolution. This is more evident in Cases II and III, since the beginning of the vortex-bubble interaction in Case I is close to the start of the recording time. After this rapid increase, the total volumetric enstrophy decreases at a rate larger than the base case, i.e. the single vortex ring \citep[see][]{foronda2021deformation}. The overall reduction in Case I highlights a loss of strength consistent with large enstrophy reductions seen in moderate-Reynolds and low-Weber-number cases, even in brief interactions~\citep{martinez2015bubbles}.  

\begin{figure}
    \centering
    \includegraphics[width=\linewidth]{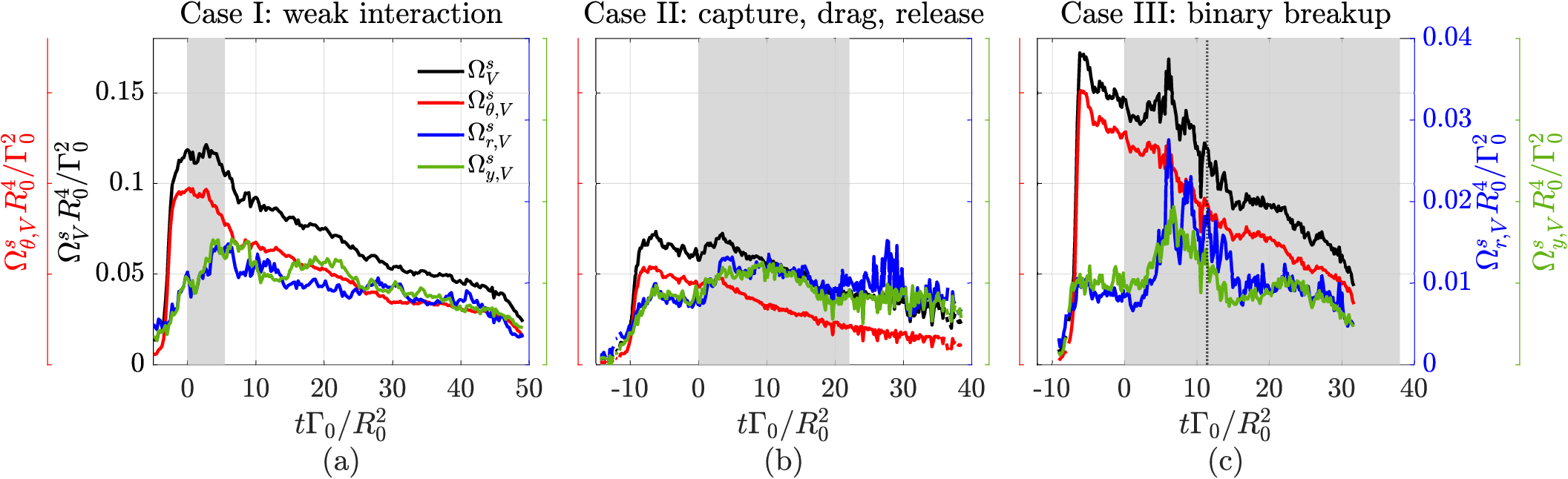}
    \caption{Specific volumetric enstrophy contributions of each vorticity component, together with the specific, total, volumetric enstrophy for (a) Case I, (b) Case II, and (c) Case III. Note that a different scale is used for the global and azimuthal enstrophy (left), and the radial and axial ones (right).}
    \label{fig:compare_enstrophy}
\end{figure}

Finally, since we are performing 3D experiments, we are able to characterise not only the contribution of the azimuthal component of the vorticity to the enstrophy, $\Omega^s_{\theta, V}$ but also the radial and axial contributions, $\Omega^s_{r, V}$ and $\Omega^s_{y, V}$, respectively. Indeed, as shown in figure~\ref{fig:compare_enstrophy}, the azimuthal contribution is the major one to the total enstrophy, being the radial and axial ones less relevant. As commented before, the volumetric enstrophy is the vortex ring property most strongly affected by the interaction, providing a global measure of the dissipation rate of kinetic energy and reflecting the overall coherence of the vortex ring. In fact, the kinetic energy of the flow is lost mainly through the creation of small-scale vorticity, that is, through an increase of enstrophy. The analysis of the contribution of the three vorticity components is crucial for understanding the three-dimensional effects of the vortex-bubble interaction. 

In Case I, an increase in both $\Omega^s_V$ and $\Omega^s_{\theta,V}$ can be observed during the vortex-bubble interaction time, which then drops sharply afterwards. This drop $\Omega^s_{\theta,V}$ is accompanied by a slight increase in $\Omega^s_{r,V}$ and $\Omega^s_{y,V}$, indicating that there is a redistribution of low-scale vorticity due to the three-dimensionality introduced by the presence of the bubble. Nevertheless, the contributions of the radial and axial vorticity components to the total enstrophy are much lower than that of the azimuthal component, and they are unable to compensate for the loss of $\Omega^s_{V}$. This is probably due to the limitations in the spatial resolution of the three-dimensional measurements, unable to resolve the smallest vorticity scales. In Cases II and III, there is also an increase of enstrophy during the interaction time, more evident in $\Omega^s_V$ than in $\Omega^s_{\theta,V}$ due to the contributions of $\Omega^s_{r,V}$ and $\Omega^s_{y,V}$. In these cases, the deformation and/or breakup of the bubble contributes to the destabilisation of the vortex ring by generating small-scale radial and axial vorticity, which increases $\Omega^s_{r,V}$ and $\Omega^s_{y,V}$, and consequently the total enstrophy, $\Omega^s_{V}$. This fact, which is especially remarkable in Case III, aligns with the results from three-dimensional numerical simulations~\citep{foronda2021deformation}. 

\FloatBarrier
\section{Conclusions}\label{sec:conclusions}
We present an experimental study on the interaction of a bubble and a vortex ring in different scenarios, corresponding to different values of the governing parameters, $Re,~We,~a_0/R_0$ and $R_0/R_b$. An experimental three-dimensional approach was chosen to characterise the volumetric nature of the problem and track bubble motion within the field of view. Specifically, we used the 4D-LPT technique based on the Shake-The-Box algorithm to obtain the liquid 3D velocity field, and we reconstructed the bubble with a tomographic algorithm (MxLOS). We applied these techniques to 4-camera and 6-camera sets of images, respectively, providing multiple points of view for an accurate inspection of the volume. We analysed the obtained velocity fields using several vortex identification algorithms available in the literature and calculated the relevant flow variables, i.e., the position and size of the vortex, and the circulation, kinetic energy, and enstrophy in the flow. Additionally, we distinguished two ways of expressing energy and enstrophy: by their classical volumetric definition or with their 2D counterparts, computed on a planar domain comprising an azimuthal cut of the volume. The latter is the equivalent of the information that could be obtained from a 2D-PIV experiment.

We compared three cases of counter-flow collision between a bubble of given size and different vortex rings, featuring the most common behaviours we found within our experiments: Case I, depicting a weak interaction, where neither the bubble nor the vortex were strongly affected; Case II, where the bubble was captured and dragged by the vortex and the latter was strongly distorted due to the presence of the bubble inside its core; and Case III, involving a stronger vortex able to capture the bubble and break it into two pieces without a severe loss of energy in the core.

The bubble-vortex interaction in Case I is brief and non-disruptive. Flow variables showed a continuous, gradual decay due to viscous diffusion, after a sharper decay due to the bubble interaction. The effect was highly localised around the bubble position, failing to spread significantly along the vortex circumference. This fact complicates addressing this type of interaction through 2D-PIV. The 3D enstrophy behaves as an average between 2D half-planes with and without the bubble. In Case II, with higher $Re$ and $We$, the bubble is captured, dragged for a finite time, and then released. The interaction caused the vortex structure to be strongly distorted. Flow variables, particularly kinetic energy and enstrophy, show a steep decay starting earlier, and their effects extend more globally around the azimuthal direction, indicating a stronger, sustained interaction. The rate of decay of the volumetric kinetic energy was larger during the interaction, and volumetric enstrophy shows an increase at the beginning of the interaction. In Case III, the combination of sufficient inertial stress (high $Re$) and a thinner core allowed the vortex to trap the bubble and stretch it azimuthally, leading to fragmentation into two daughter bubbles. This case behaves very similarly to Case II, showing significant dynamics in enstrophy: it concentrated heavily in the bubble area, peaked just before breakup (related to azimuthal elongation), and subsequently dropped quickly. The 3D enstrophy here revealed an interplay between global decay and near-bubble local effects. Furthermore, all cases have depicted an increase of the non-azimuthal contributions of enstrophy during the interaction, pointing to a redistribution of vorticity due to the presence of the bubble.

Note that the primary difference between Cases II and III lies in how the vortex core thickness and the resulting stress distribution affect both the stability of the vortex ring and the bubble integrity. Case II, characterised by a thicker core and a higher Weber number ($We = 1$), leads to bubble capture and dragging, which induces strong structural deformation and a loss of axial symmetry in the vortex ring. In this case, both kinetic energy and azimuthal enstrophy exhibit a steep, early decay that extends globally in the azimuthal direction, indicating a sustained and highly disruptive interaction. Conversely, Case III involves a thinner vortex core that concentrates stresses more intensely, enabling the vortex to trap, azimuthally stretch, and eventually break the bubble, while the ring itself remains largely coherent. Although both cases show an initial increase in volumetric enstrophy at the onset of the interaction, Case III is distinguished by a localised enstrophy peak within the bubble just before fragmentation, associated with bubble elongation, followed by a redistribution of vorticity towards the daughter bubbles. Finally, the rate of kinetic energy decay in Case III accelerates significantly after breakup, whereas in Case II the most substantial vorticity loss occurs after the bubble is released.

By employing advanced 3D diagnostics, this work addresses the challenge of capturing the complex, transient, and non-linear phenomena, providing valuable, high-fidelity data to address questions about the actual 3D evolution of vorticity and enstrophy during bubble-vortex interactions. The detailed characterisation of the bubble-vortex interaction yields crucial insights into the physical mechanisms governing the coupling between the phases. Future work could fruitfully extend this analysis with the detailed temporal evolution of the bubble dynamics—directly accessible from the reconstruction data—aiming to clarify the fundamental mechanisms driving bubble deformation and eventual breakup. Furthermore, we plan to complement the flow analysis with data assimilation techniques, which will provide further resolution of the flow around the bubble, a key step to estimate accurately the turbulent stresses and the pressure field during the interaction process. These findings are essential for advancing physics-based models of bubble breakup and turbulence modulation in complex bubbly flows.

\backsection[Supplementary data]{\label{SupMat}Supplementary movies are available at https://doi.org/...}

\backsection[Acknowledgements]{ The authors acknowledge the CNRS Research Federation FERMaT (FR3089) for giving access to the 4D-LPT system at IMFT and thank H. Ayroles, C. Perez and L. Mouneix for the technical support provided for the experimental setup at IMFT.}

\backsection[Funding]{This work was partly supported by projects PID2023-151343NB-C32 and PID2022-140433NA-I00, financed by MCIN/AEI/10.13039/501100011033/ FEDER, UE. J.R.R. acknowledges the financial support provided by the European Union through the NextGenerationEU Fellowship UJAR06MS. R.B.J would like to acknowledge the project R1C\_2025\_011 funded by the Research and Knowledge Transfer Internal Plan of the University of Ja\'en. C.E.C. would like to thank the Spanish Ministry of Universities for the financial support provided by the Fellowship FPU20/02197.  M.L.D. also acknowledges the grant RYC2023-044496-I financed by MICIU/AEI /10.13039/501100011033 and FSE+. P.E. and S.C. acknowledge financial support from Toulouse INP, international program ETI.}

\backsection[Declaration of interests]{The authors report no conflict of interest.}

\backsection[Data availability statement]{The data that support the findings of this study are available upon reasonable request}

\appendix

\section{Calibration procedure}\label{app:calib}
\begin{figure}
    \centering
    \includegraphics[width=\linewidth]{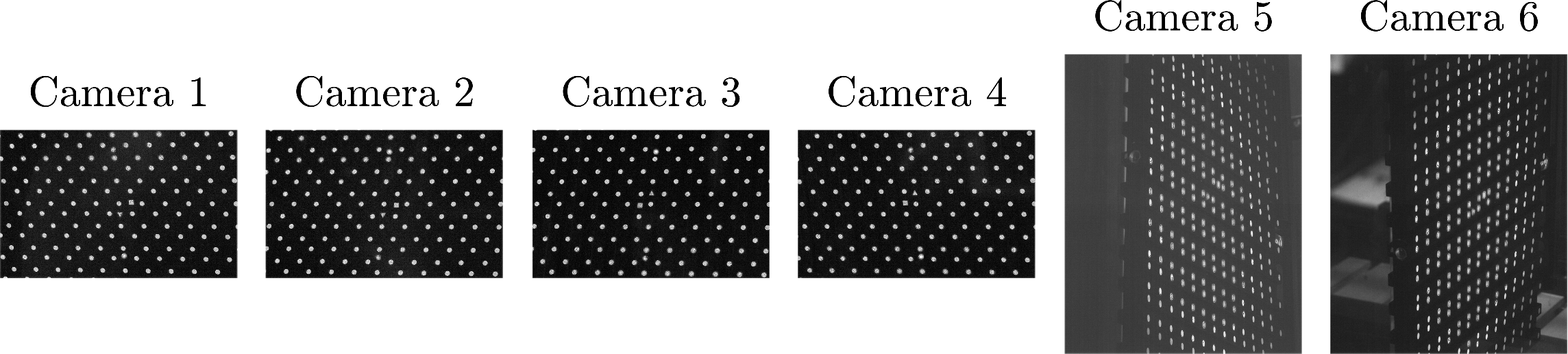}
    \caption{Perspective views of the calibration target from each camera.}
    \label{fig:calibration}
\end{figure}

A bi-planar two-faces calibration target from LaVision GmbH (Ref. No. 204-15) was immersed in the tank filled with water and placed opposite the front cameras. To cover the full volume of interest, the target was recorded in multiple positions (11 positions 10 mm apart), assuring that its centre could still be seen from all the cameras (see figure \ref{fig:calibration}). Given the complex camera arrangement, the lateral view of the calibration plate was only partial, so the lateral images were masked to guarantee that only the plate pattern was taken into account for the calibration.   
The pattern was manually interpreted with DaVis software, which then computed the 3D calibration function. This function consisted of a polynomial fit which provided a mapping from the camera sensor coordinates to the real-world ones for each camera, while considering potential refraction due to the light crossing different media in between~\citep{pavlov2021exploration}. The calibration was corrected for every experimental set by performing a volumetric self-calibration of uncorrelated images of the particles, which iteratively modifies the original calibration function to reduce disparities accounting for many sources of errors, such as optical distortions, vibrations, thermal expansions, or experimental uncertainties on the calibration process \citep{wieneke2008volume,wieneke2018improvements}. The resulting optical transfer function (OTF) considers the shape of each detected particle, which enables easier identification of such a particle in the successive images, reducing the probability of ghost particles \citep{schanz2012non}. As a result of this complex procedure, the images are scaled to 17.91 px/mm.
The particle images were pre-processed to condition the particle intensity and to reduce noise, so that the Shake-The-Box (STB) technique \citep{schanz2016shake} may be employed to capture the evolution of the flow.

\section{Characterization of the flow field: 4D-LPT}\label{app:STB}
The liquid velocity field was characterised using 4D Lagrangian Particle Tracking (4D-LPT), a laser velocimetry technique based on the Shake-The-Box algorithm implemented in the DaVis LaVision software, which allowed us to reconstruct three-di\-men\-sional tracer particle trajectories over time, providing volumetric velocity fields and enabling the analysis of complex, unsteady flow structures.

The Shake-The-Box (STB) technique \citep{schanz2016shake} was employed to capture the evolution of the flow field using the four frontal cameras. The results using six cameras were similar qualitatively, but fewer particles could be tracked as the thickness of the measurement volume seen by the two lateral cameras is larger than their depth of field, the image quality being thus affected by the amount of out-of-focus highlighted particles. Once we performed an accurate calibration of the 4 frontal cameras (see Appendix \ref{app:calib}), the STB technique is able to determine the trajectories of the particles resolved in time in 3D at very high particle densities, greater than 0.05 particles per pixel~\citep{schanz2013shake,schanz2016shake}. This is possible because of the predictive nature of this algorithm, which uses the information of existing particle tracks to anticipate the most probable optical solution for the following time step instead of reconstructing the whole volume, thus reducing ghost particles and computational resources \citep{schanz2016shake,lorite2022experimental}. This initial guess is then refined to match the recorded images.

In our experiments, the images were preprocessed to maximise the contrast between the particles and the background. However, to avoid possible disturbances due to reflections, a threshold was set on image intensity, so that only pixels above 12-20\% of the maximum intensity were considered as possible particles. Additionally, velocity and acceleration thresholds were manually set according to the apparent maximum displacement of the particles in the area of interest (the maximum velocities were expected at the centre of the ring and at the cores) and were then adjusted after several iterations so that the maximum velocity could be captured with the minimum particle identification errors. To further refine the reconstruction of the paths, we set a minimum track length of 5 time steps to neglect any possible ghost particles, and included several iterations where we gradually decreased the temporal increment (from 6 to 1 time step), alternating passes both forward and backwards in time, while reconnecting interrupted paths between passes. With this method and an allowed search volume of 1.5 voxels, we achieved around 80 to 100 thousand tracks, depending on the specific seeding conditions in each experiment. The velocities of the particles were derived by fitting the trajectories with a polynomial of order 3 to compute the tracks for each time step, considering a time window of 15 steps. Finally, the Lagrangian trajectory data were converted to an Eulerian grid, providing cubic sub-volumes containing a number of particles representative enough so that the resulting vectors could describe the fluid motion accurately (typical grid size between 13 and 18 voxels, implying a resolution between 0.7 and 1 mm). With respect to the velocity measurements, considering that the relative uncertainty scales with the local particle displacement, we estimate an error below 2\% in high-velocity regions (such as the vortex core) and up to a conservative 10\% in the far-field regions where displacements are minimal~\citep{sciacchitano2021main}.
\\

\section{Bubble shape and motion: tomographic reconstruction}\label{app:bubble}
\begin{figure}
    \centering
    \includegraphics[width=\linewidth]{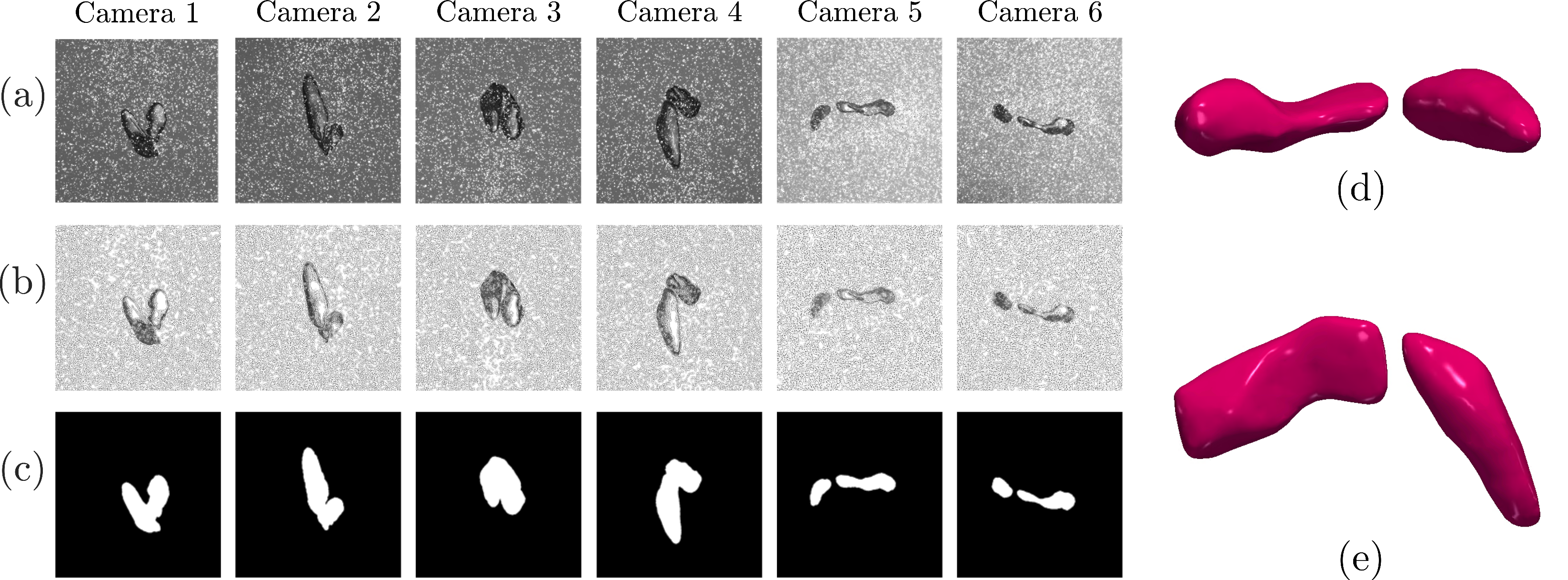}
    \caption{Left: snapshots of a broken bubble, as viewed from each camera: (a) raw images, (b) preprocessed images, (c) isolated objects. Right: volumetric reconstruction of the bubble, (d) $yz$ view, (e) $xz$ view. A supplementary movie related to this figure (Movie1\_reconstruction.mp4) is accessible at journals.cambridge.org.}
    \label{fig:reconstruction}
\end{figure}

To analyse the bubble dynamics, the shadow projections of the bubble on each camera were isolated from the greyscale images of the six cameras (figure~\ref{fig:reconstruction}a) by means of a segmentation procedure similar to that by \cite{boulesteix2010cisaillement}.
Firstly, the background was calculated by means of a median filter, and the image was divided over the background to enhance the contrast of the bubble (see figure~\ref{fig:reconstruction}b). Then, a hysteresis threshold was applied to the gradient image and the contour of the bubble was isolated by finding the local maxima in the gradient image in the direction perpendicular to the gradient. Finally, the remaining open contours were closed by filling the image in the direction of the gradient and using consecutive dilation and erosion operations. The isolated bubbles are shown in figure~\ref{fig:reconstruction}(c).

The commercial software DaVis 11 from LaVision GmbH was used to reconstruct a  three-di\-men\-sional volume using the six available bubble projections and an accurate 3D calibration (see Appendix \ref{app:calib}). The camera arrangement was essential for the quality of the tomographic reconstruction. On the one hand, the small angles between each pair of front cameras allow for an accurate determination of the position of the bubbles in a vertical plane. On the other hand, the large angles between the front and lateral cameras provide a variety of points of view, which helps to constrain the reconstruction, avoiding the distortion of the volume due to a lack of depth perception \citep{todter2025application}.

The Multiplicative Line‑Of‑Sight (MLOS) reconstruction technique was used for the reconstruction of the bubble. The resulting objects, shown in figure~\ref{fig:reconstruction}(d), (e), were then corrected by eliminating the potential ghost volumes in the areas covered by the bubbles in the lines of sight of the cameras. After the objects were reconstructed, the volumes, trajectories, and velocities of the bubbles could be obtained. Excellent agreement was obtained between the reconstructed bubble volume and the axisymmetric volume inferred from the additional camera images at the end of the bubble formation process, even in the presence of bubble breakup events.\\

\section{Data processing}\label{app:processing}

\begin{figure}
    \centering
    \includegraphics[width=\linewidth]{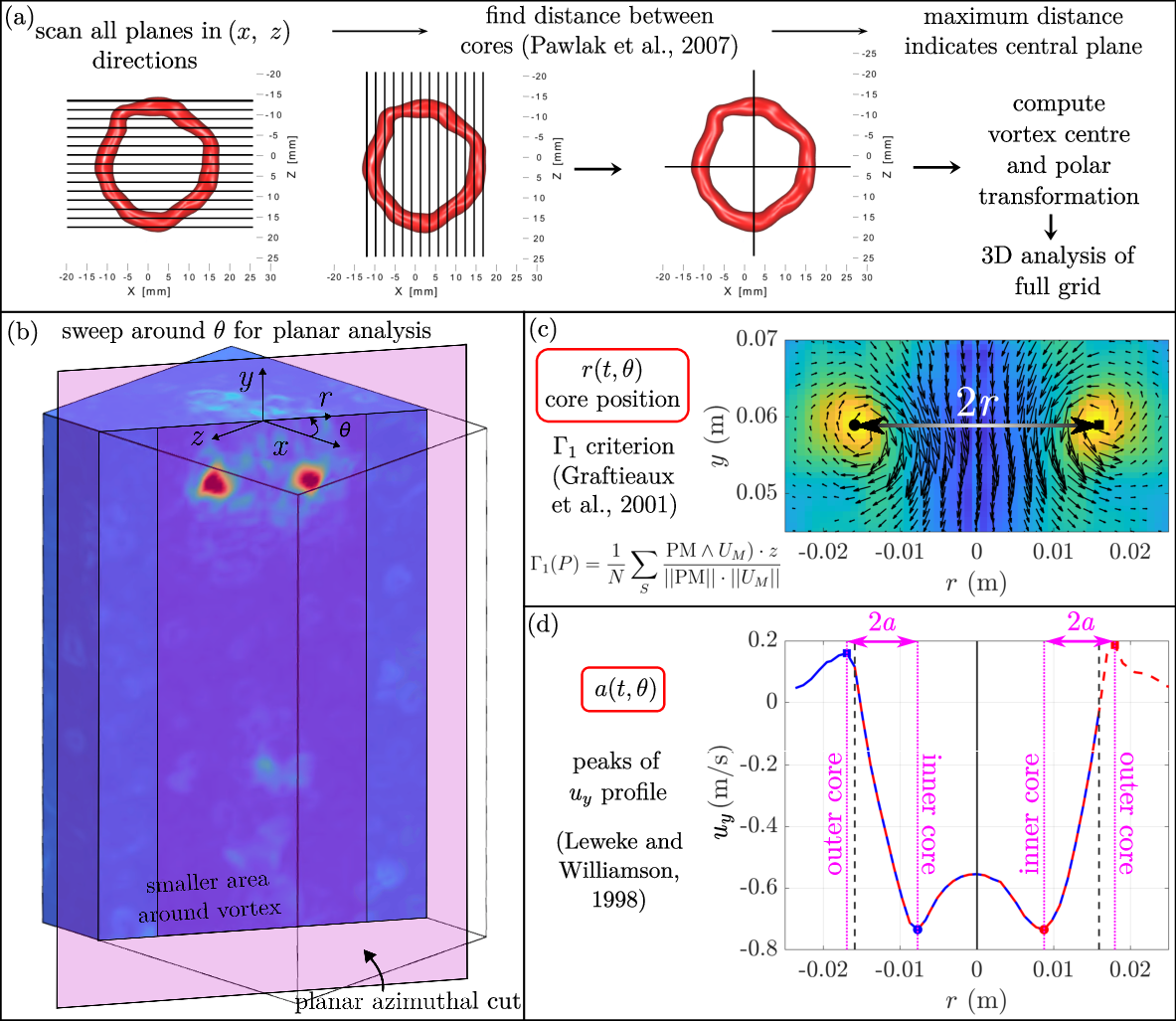}
    \caption{Block diagram of the post-processing routine used to locate the vortex, calculate its size and compute all its characteristic variables. In (c), $\Gamma_1$ is a scalar function used to identify the vortex ring. $S$ is a rectangular domain of fixed size and geometry, centred in P, and $N$ is the number of points M inside $S$. PM is the vector between P and each of the points M in the grid, and $U_M$ is the velocity vector at M.}
    \label{fig:block_diagram}
\end{figure}
The procedure followed to extract the vortex ring characteristics from the 3D velocity field is sketched in figure~\ref{fig:block_diagram}, and is described as follows. The first stage was to locate the vortex ring within the 3D velocity field for each time step. 
Our approach was to deal with the 3D data as a set of consecutive 2D planes and to apply vortex identification algorithms for 2D-PIV data that are available in the literature. To do this, we would sweep along one of the horizontal directions (e.g. $z$) and obtain consecutive planar velocity fields (one $xy$ plane at every $z$ coordinate of the 3D grid), see figure~\ref{fig:block_diagram}(a). On each individual plane, we applied a set of dimensionless integral functions proposed by \cite{pawlak2007experimental} to identify the vortex position. From the integration of the radial velocity along the axial coordinate \citep[see equation~(7) in][]{pawlak2007experimental}, we obtain the vertical position of the two vortex cores present in each plane ($y_{cl},~y_{cr}$, where $cl$ stands for left core and $cr$, for right core). Equation (9) in \citet{pawlak2007experimental}, responsive to variations in the momentum fluxes across the surface of a cylinder aligned with the $y$ axis, was applied at $y_{cl}$ and $y_{cr}$, giving their horizontal coordinates (e.g. $x_{cl},~x_{cr}$). The diameter of the vortex was calculated as the distance between the cores: $d_z=\sqrt{(y_{cl}-y_{cr})^2+(x_{cl}-x_{cr})^2}$. The plane where the diameter of the vortex was maximum marked the coordinate of the centre of the vortex ring in such direction, $z_v$. The same procedure was repeated for the remaining horizontal direction (e.g. sweep along $x$ to find $x_v$). The crossing point between the central planes ($x_v,~z_v$) will define the centre of the vortex at each time step (see figure~\ref{fig:block_diagram}a), along with the vertical coordinate, $y_v$, defined as the average of the vertical positions of the cores at the central plane in each direction ($y_{cl_x},~y_{cr_x},~y_{cl_z},~y_{cr_z}$). The difference between these vertical coordinates was generally under 2 mm.

The next step was to compute a polar transformation of the coordinates to capture the cylindrical-symmetric nature of the problem, namely ($x,z,y)\rightarrow (r,\theta,y$). Note that the origin of the reference frame was located at ($x_{v_0},z_{v_0},y_{v_0}$), i.e., the coordinates of the vortex ring centre at the initial time step, $t=t_0$. Similarly to the previous procedure, the volume was sectioned, but this time, azimuthally, i.e., we selected one $yr$ plane for each $\theta$ value, and each division was analysed individually in 2D. At first approach, these vertical planes cover both cores and thus $\theta$ will vary between 0 and 180$^\circ$ (full planes hereinafter). The volume was sectioned into 80 divisions containing two cores, thus providing 160 sections (half-planes hereinafter) where the behaviour of the vortex ring core could be analysed. Therefore, each one of the 80 planes yields a planar velocity field, similar to those obtained by 2D-PIV measurements. In this way, for an ideal, unperturbed vortex ring, the information in all the divisions would be the same: two cores of radius $a$ separated horizontally by a distance $2R$ (see figure~\ref{fig:block_diagram}b). Nevertheless, this picture is expected to change in the presence of a bubble interacting with the vortex ring. By analysing several $\theta$ positions instead of a fixed one, this approach allows us to define the precise circular sectors where the bubble is present ($\theta_{bubble}$ in figure~\ref{fig:problem_sketch}b) and whether the effect of the interaction is local at $\theta_{bubble}$, it moves azimuthally with the bubble, or it spreads around $\theta$ within the vortex core. 

Once we have the azimuthal cuts of the flow field, the precise vortex core centres are identified and located based on a scalar function, $\Gamma_1$, which was derived from the topology of the velocity field \citep[see Eqs. 8 and 9 in][]{graftieaux2001combining}. When applied on a $yr$ plane, the function provided peaks at the vortex cores present on such plane (see figure~\ref{fig:block_diagram}c), from where the coordinates of each peak could be obtained ($y_{cl},~r_{cl}$ on the left, and $y_{cr},~r_{cr}$ on the right). The differences between the radial coordinate of the centre of the vortex and its cores give us the vortex radius at each azimuthal position: $r(\theta)=r_{cr}-r_v,~r(\theta+180^\circ)=r_{cl}-r_v$, with $r_v=\sqrt{x_v^2+z_v^2}$ (see black dashed lines in figure~\ref{fig:block_diagram}d). Note that $r_v$ indicates the instantaneous radial position of the centre of the vortex, which may deviate from the $y$-axis.

Having determined the position of the vortex cores in each $\theta$-plane, we may also compute their size. Within the azimuthal planar domain, we analysed the vertical velocity profile on a horizontal line crossing each core: $u_y(y_{cl},~r_l)$, with $r_l$ the radial domain of the left half-plane, resp. $u_y(y_{cr},~r_r)$ for the right side. These profiles exhibit a positive and a negative peak at the outer and inner end of the core, respectively \citep{sullivan1973study,leweke1998cooperative}, as shown in figure~\ref{fig:block_diagram}(d), where the red line is the velocity profile on the right core, and the blue line is the one on the left core. The radial distance between such peaks provided the core diameter, $2a$.

\bibliographystyle{jfm}
\bibliography{jfm}


\end{document}